\documentclass[sigconf]{acmart}
\renewcommand\footnotetextcopyrightpermission[1]{} % removes footnote with conference information in first column abstract

\usepackage{listings}
\usepackage{colortbl}
\usepackage{algorithm}
\usepackage{algpseudocode}
\usepackage{hyperref}
\usepackage{hyperxmp}

\AtBeginDocument{%
  \providecommand\BibTeX{{%
    \normalfont B\kern-0.5em{\scshape i\kern-0.25em b}\kern-0.8em\TeX}}}

\copyrightyear{2025}
\acmYear{2025}
\setcopyright{cc}
\acmConference[CHI '25]{CHI Conference on Human Factors in Computing Systems}{April 26-May 1, 2025}{Yokohama, Japan}
\acmBooktitle{CHI Conference on Human Factors in Computing Systems (CHI '25), April 26-May 1, 2025, Yokohama, Japan}\acmDOI{10.1145/3706598.3713740}
\acmISBN{979-8-4007-1394-1/25/04}
\newif\ifsubmit
\submitfalse
\ifsubmit
    \newcommand{\stefanie}[1]{}
    \newcommand{\mackenzie}[1]{}
    \newcommand{\faraz}[1]{}
    \newcommand{\liane}[1]{}
    \newcommand{\changes}[1]{}
    \newcommand{\donald}[1]{}

\else
    \newcommand{\stefanie}[1]{{\leavevmode\color[rgb]{1.0, 0.0, 0.5}{#1}}}
    \newcommand{\mackenzie}[1]{{\leavevmode\color[rgb]{1.0, 0.6, 0.0}{#1}}}
    \newcommand{\faraz}[1]{{\leavevmode\color[rgb]{0., 0, 0}{#1}}}
    \newcommand{\changes}[1]{{\leavevmode\color[rgb]{0.0, 0.0, 0.0}{#1}}}
    \newcommand{\donald}[1]{{\leavevmode\color[rgb]{0.9, 0.4, 0.5}{#1}}}
    \newcommand{\liane}[1]{{\leavevmode\color[rgb]{0, 0.7, 1.0}{#1}}}

\fi

\usepackage{soul}

\usepackage{booktabs}
\usepackage{multirow}
\usepackage{graphicx}
\usepackage{microtype}

\begin{document}

% \title[TactStyle]{TactStyle: Enhancing 3D Model Stylization with Realistic Texture Generation for Digital Fabrication}

\title[TactStyle]{TactStyle: Generating Tactile Textures with Generative AI for Digital Fabrication}

% \title{Enhancing 3D Model Stylization with Tactile Texture Generation for Digital Fabrication} 

\author{Faraz Faruqi}
% \authornote{Both authors contributed equally to this research.}
\email{ffaruqi@mit.edu}
\orcid{0000-0002-1691-2093}
% \author{G.K.M. Tobin}
% \authornotemark[1]
\affiliation{%
  \institution{MIT CSAIL}
  \city{Cambridge}
  \state{MA}
  \country{USA}
  % \postcode{43017-6221}
}

\author{Maxine Perroni-Scharf}
% \authornote{Both authors contributed equally to this research.}
\email{max1@mit.edu}
\orcid{0000-0002-4075-5745}
% \author{G.K.M. Tobin}
% \authornotemark[1]
\affiliation{%
  \institution{MIT CSAIL}
  \city{Cambridge}
  \state{MA}
  \country{USA}
  % \postcode{43017-6221}
}

\author{Jaskaran Singh Walia}
% \authornote{Both authors contributed equally to this research.}
\email{karanwalia2k3@gmail.com}
\orcid{0000-0002-9255-5446}
% \author{G.K.M. Tobin}
% \authornotemark[1]
\affiliation{%
  \institution{Vellore Institute of Technology}
  \city{Chennai}
  % \state{MA}
  \country{India}
  % \postcode{43017-6221}
}

\author{Yunyi Zhu}
% \authornote{Both authors contributed equally to this research.}
\email{yunyizhu@mit.edu}
\orcid{0000-0003-4545-8069}
% \author{G.K.M. Tobin}
% \authornotemark[1]
\affiliation{%
  \institution{MIT CSAIL}
  \city{Cambridge}
  \state{MA}
  \country{USA}
  % \postcode{43017-6221}
}

\author{Shuyue Feng}
% \authornote{Both authors contributed equally to this research.}
\email{shuyuefeng@zju.edu.cn}
\orcid{0000-0002-6720-5356}
% \author{G.K.M. Tobin}
% \authornotemark[1]
\affiliation{%
  \institution{Zhejiang University}
  \city{Hangzhou}
  % \state{MA}
  \country{China}
  % \postcode{43017-6221}
}

\author{Donald Degraen}
\email{donald.degraen@canterbury.ac.nz}
\orcid{0000-0003-1029-931X}
\affiliation{%
  % \department{HIT Lab NZ}
  \institution{HIT Lab NZ, University of Canterbury}
  \streetaddress{Private Bag 4800}
  \city{Christchurch}
  \country{New Zealand}
  \postcode{8140}
}

\authornote{
Also with HCI Group, University of Duisburg-Essen.
}
% \additionalaffiliation{%
%   \institution{University of Duisburg-Essen}
%   \streetaddress{Schützenbahn 70}
%   \city{Essen}
%   \country{Germany}
%   \postcode{45127}
% }
% \affiliation{%
%   \institution{University of Duisburg-Essen}
%   \streetaddress{Schützenbahn 70}
%   \city{Essen}
%   \country{Germany}
%   \postcode{45127}
% }

\author{Stefanie Mueller}
\email{stefanie.mueller@mit.edu}
\orcid{0000-0001-7743-7807}
% \author{G.K.M. Tobin}
% \authornotemark[1]
\affiliation{%
  \institution{MIT CSAIL}
  \city{Cambridge}
  \state{MA}
  \country{USA}
  % \postcode{43017-6221}
}

\renewcommand{\shortauthors}{Faruqi et al.}

% Main potential reviewers:
% Jeeeun Kim @ HCIED (Did design tool for softness by example)
% Michal Piovarci (Does perceptual modeling for fabrication)
% Roland Bennwitz (Does material fabrication and tactile perception)
% Ryo Suzuki (Did surface texture generation in Tabby)
% Kentaro Yasu (Did surface textures through magnetic feedback)
% Haruki Takahashi @ Meiji University (Did surface texture fabrication and haptic fabrication)

% Others:
% Cesar Torres (Did surface texture fabrication)
% Xin Yan @SDU (Does geometric modeling and fabrication)
% Chase Timms @ Meta (Did perceptual modeling of tactile roughness)
% Riad Sahli (Did perceptual studies of fabricated structures)

\begin{abstract}

Recent work in Generative AI enables the stylization of 3D models based on image prompts. However, these methods do not incorporate tactile information, leading to designs that lack the expected tactile properties. We present TactStyle, a system that allows creators to stylize 3D models with images while incorporating the expected tactile properties. TactStyle accomplishes this using a modified image-generation model fine-tuned to generate heightfields for given surface textures. By optimizing 3D model surfaces to embody a generated texture, TactStyle creates models that match the desired style and replicate the tactile experience. We utilize a large-scale dataset of textures to train our texture generation model. In a psychophysical experiment, we evaluate the tactile qualities of a set of 3D-printed original textures and TactStyle's generated textures. Our results show that TactStyle successfully generates a wide range of tactile features from a single image input, enabling a novel approach to haptic design.

% Finally, we present TactStyle's user interface, which allows creators to personalize 3D models with desired visual and tactile properties.

 \end{abstract}



%% the work being presented. Separate the keywords with commas.

\begin{teaserfigure}
\centering
  \includegraphics[width=\textwidth]{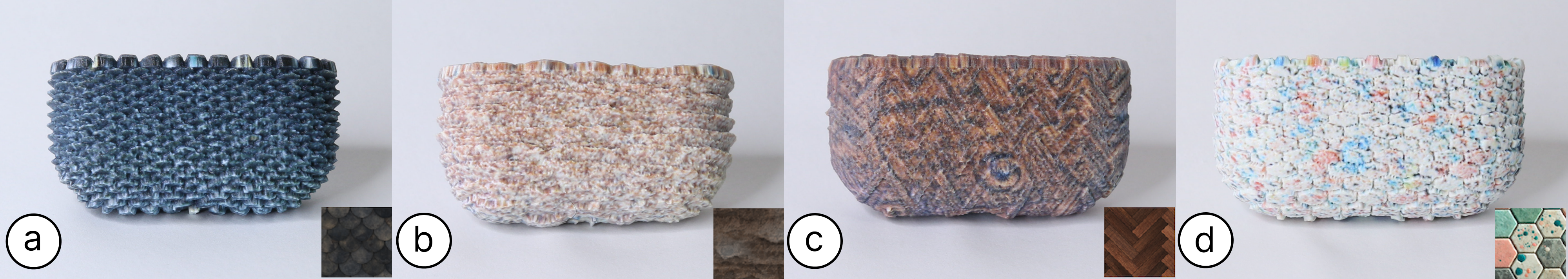}
  \vspace{-12pt}
  \caption{TactStyle allows creators to stylize 3D models with image input while incorporating the tactile properties of the texture in addition to its color. Here, we show four different textures applied to the same 3D model, an Airpods cover, with the image stylization prompt shown on the bottom right. The different textures used are: a) round stone roof, b) layered brown rock, c) herringbone wood, and d) colorful hexagonal tiles.}
  \label{fig:teaser}
\end{teaserfigure}

%%
%% This command processes the author and affiliation and title
%% information and builds the first part of the formatted document.
\maketitle

\section{Introduction}

%\faraz{add relevant papers from fabrication}

With the growing popularity of 3D printing research within HCI~\cite{mueller20173d}, there is also increasing interest in developing tools that enable users to customize 3D models. Open-source repositories, such as Thingiverse~\cite{Thingiverse2024Apr}, are a useful resource for ready-to-print 3D~models. However, their customization is limited to changing predefined parameters~\cite{alcock2016barriers}. Recent advances in Generative AI allow users to more freely customize their 3D models using text prompts or images as user-provided style descriptions~\cite{faruqi2023style2fab, michel2022text2mesh}. However, these existing frameworks for stylizing 3D models primarily focus on modifying the model to match a desired visual appearance as described via the provided text or image prompt~\cite{piovarci2020towards}.  %Visual  and auditory [CITE] modalities have also been extensively explored in the design and fabrication of interactive systems.

% When approximating the visual appearance, existing Generative AI stylization frameworks, such as Text2Mesh [ref], modify the surface texture of the model. This leads to unexpected tactile properties 

One underexplored area of model customization is texture—specifically, the `tactile feedback' of printed structures, such as whether a surface feels smooth, rough, or potentially reminiscent of materials like wood grain or stone. Our ability to sense these textures through touch plays a crucial role in our interactions with the physical world, shaping not only how we perceive and manipulate objects but also influencing our emotional and cognitive responses to them~\cite{gallace2014touch}.  Therefore, augmenting printed structures with appropriate tactile properties can enrich interaction with physical objects, especially when mimicking materials that differ from the 3D printing material. 

Recent advances in computer vision have proposed methods for capturing high-fidelity visual properties from images, enabling digital replication of textures from real-world surfaces \cite{yeh2024texturedreamer, huang2023nerf}. However, these techniques are limited to digital replication, as highlighted in TextureDreamer~\cite{yeh2024texturedreamer}, where they do not optimize normal maps to avoid details that are inconsistent with the target mesh. In the field of digital fabrication, researchers have proposed techniques to capture surface microgeometry~\cite{degraen2021capturing} as a heightfield and use this data to replicate the texture using fabrication methods such as 3D printing. However, they currently require sophisticated equipment such as photometric sensing techniques~\cite{johnson2009retrographic} to capture the surface microgeometry of each texture, limiting their usability. Thus, image-based replication is currently limited to replicating visual elements of textures, and creators are currently limited to replicating textures through associated digital surface microgeometry data. We hypothesize that by learning a correlation between a texture's visual image, and its heightfield (surface microgeometry), we can replicate the tactile properties of textures, directly from an input image.  

We present TactStyle, a system that allows creators to stylize 3D models with texture images while incorporating the expected tactile properties. TactStyle accomplishes this by separating the visual and the geometry stylization, and augmenting the process with a novel geometry stylization module that replicates the tactile properties of textures based on user input. The novel geometry stylization module uses a fine-tuned variational autoencoder (VAE)~\cite{kingma2013auto}, that translates the user provided visual image of a texture into a surface microgeometry or heightfield. The model then uses this heightfield to manipulate the geometry to create the tactile properties on the 3D model surface. Separately, the visual appearance of the 3D model is optimized with a method~\cite{faruqi2023style2fab} that has been shown to accurately replicate visual qualities. Thus, by optimizing 3D model surfaces to embody both the color and tactile properties of a given input image, TactStyle allows creators to generate stylized models that not only visually match the desired style but also replicate the tactile experience.

\section{Related Work}
To situate our question and findings, we draw upon previous research in personalizing open-source designs, 3D printed haptics, and tactile surface reconstruction to develop our proposed system.

\subsection{Personalizing Open-Source Designs}
As 3D printing has become the preferred digital fabrication tool for both expert and amateur makers alike~\cite{Kuznetsov_2010_expertamateur, faruqi2021slicehub}, makers have increasingly shared their 3D models and designs through online open repositories, such as Thingiverse. Alcock et al.~\cite{alcock2016barriers} propose that such repositories serve as ideal training platforms for novice makers. However, several challenges remain in allowing creators to personalized open-source designs. Numerous studies have investigated novice makers~\cite{hudson2016understanding, norouzi2021making} in the process of modifying shared 3D models. These studies consistently find that users face significant challenges when attempting to modify existing designs. Editing 3D printable meshes requires advanced expertise of computer-aided design workflows~\cite{schmidt2010meshmixer} which are often incompatible with the skill levels of novice users. Moreover, as highlighted by Oehlberg et al.~\cite{oehlberg2015patterns}, even when designs are customizable, they often fall short of the scope of modifications makers desire. For instance, customizing 3D models to support varying tactile properties is rarely supported.

\subsection{3D Printed Haptics \& Tactile Surfaces}

3D printing haptics has been an increasing area of interest in HCI. Design tools have been developed that provide users with the ability to customize their designs for a variety of haptic properties, ranging from desired heat dissipation (Thermal Comfort~\cite{thermalcomfort2017}), to customizable stiffness (X-Bridges~\cite{xbridges2022}) and softness  (OmniSoft~\cite{omnisoft2021}), and personalized force feedback (Shape-Haptics~\cite{shapehaptics2022}).

To 3D print surfaces with desired tactile properties, researchers developed a range of new 3D printing techniques. For instance, 3D printed Hair~\cite{laput20153d} creates hair-like structures and bristles through FDM 3D-printing by exploiting the stringing phenomena inherent to the 3D-printing process. Cillia~\cite{cillia2016} creates high-resolution tactile surfaces by modifying the input to the SLA printing process. Such hair-like structures have been shown to influence texture perception, especially when combined with visual augmentation~\cite{degraen2019enhancing}.
On-the-Fly Fine Texture 3D Printing~\cite{texture3Dprinting2021}, Thickness Control~\cite{thicknesscontrol2016} and ExtruderTurtle~\cite{extruderturtle2022} modify the G-code underlying FDM 3D printers to create varying types of tactile surface textures. Another approach to create 3D printed tactile surfaces is to print mechanical structures that can transform into various surface structures. For instance, Metamaterial Textures~\cite{metamaterialtextures20218} demonstrated how varying tactile surface can be created within a single 3D print, and such structural approaches have been shown to directly influence perception during tactile fingertip exploration~\cite{feick2023metareality}.
Most closely aligned with our approach is HapticPrint~\cite{torres2015hapticprint}, which modifies the ``feel'' of 3D-printed objects by providing a tool to automatically generate heightfields by grayscaling raster images to create tactile textures on arbitrary 3D geometries. This approach gives an approximate texture from the image, which can then be applied to the 3D model. However, a grayscale image and a heightfield differ since grayscale only encodes luminance, not the local height of the surface. Since not all color changes in an image are related to changes in depth or height~\cite{yeh2024texturedreamer}, taking a grayscale version would fail to distinguish between color patterns caused by surface detail and those caused by lighting. In this work, we modify an image-generation model to generate heightfield data, by fine-tuning a diffusion-based model on pairs of texture-heightfield data. 

\subsection{Reconstructing Tactile Surfaces}
% Prior work in HCI has had a longstanding focus on recreating haptics. Researchers have explored several different aspects of haptic interaction for digital fabrication. For instance, prior work has explored electronic augmentations to digital fabrication techniques for haptic properties. For instance, Liquido~\cite{schmitz2016liquido} \stefanie{embeds liquid in 3D-printed objects and uses capacitive sensing to sense tilting and motion interactions}, and Tacttoo~\cite{withana2018tacttoo}, provides a feel-through electric haptic tattoo that outputs feedback directly onto users' skin. Similarly, Tran et al.~\cite{tran2023augmenting} use vibrotactile actuation on users' fingernails to provide haptic feedback. 

Researchers have also explored augmenting a 3D model's geometry to enable accurate tactile perceptions. Haptography~\cite{kuchenbecker2011haptography} introduces an approach that uses sensors to capture the haptic properties of real objects and recreate them in virtual environments, while Metareality~\cite{feick2023metareality} designed adaptable metamaterial structures that can alter their hardness and roughness upon compression. Degraen et al~\cite{degraen2021capturing} showed how a real-world texture's tactile properties can be replicated by using its microgeometry processed with a photometric sensing technique~\cite{johnson2009retrographic}. Existing Generative AI-based stylization methods such as Style2Fab~\cite{faruqi2023style2fab} and Text2Mesh~\cite{michel2022text2mesh} allow users to stylize their 3D models based on text and image prompts. These methods perform iterative refinement of the mesh, making small changes on vertex and color channels of the 3D model and estimating its similarity to the goal text or image prompt provided by the user. However, since these methods are based on image-based losses, replicating the surface microgeometry becomes a challenge~\cite{faruqi2024shaping}.

% Bring back if needed. 
% To calculate this similarity, these methods leverage CLIP~\cite{radford2021learning}, a latent representation learned using pairs of images and their text captions. CLIP makes it possible to compute a similarity metric between image-image or text-image pairs, termed `CLIP Loss', which can be used to compare the visual appearance of a model being stylized and guide the stylization towards a better approximation. CLIP, however, lacks information about the geometrical patterns associated with each texture that is essential to replicate the tactile properties associated with it~\cite{degraen2021capturing}. 

% However, these tools do not enable creators to customize existing 3D models to replicate a specific texture's tactile properties.
% \faraz{changes here}
TactStyle extends this line of work by proposing a system that allows creators to stylize 3D models using images as input, optimizing not only the texture's appearance but also its expected tactile properties.

\section{Formative Study}

We hypothesize that current stylization frameworks that leverage \faraz{latent representations~\cite{faruqi2023style2fab} such as CLIP~\cite{radford2021learning}} are efficient at replicating the visual appearance of a texture but ineffective at replicating its tactile properties. We first test this hypothesis by performing a formative study. Although stylization frameworks allow both text and image-based stylization, we consider only image-based stylization for our experiments. This is because text-based stylization methods require a captioning technique to generate textual descriptions of textures, which may not express all its details. This limitation was also highlighted by TextureDreamer~\cite{yeh2024texturedreamer}. Thus, in this formative study, we focus on testing the stylization of a 3D model based on image prompts. 

\subsection{Dataset and Stylization Baseline}
To investigate the accuracy of texture replication, we use a large-scale dataset of PBR (Physically Based Rendering) textures from CGAxis~\cite{cgaxis_pbr_20_parquets}. This dataset contains both visual and heightfield information about textures. We collect a total of 500 textures which contain textures for `\textit{Parquets}', `\textit{Wood}', `\textit{Rocks}', `\textit{Walls}', and `\textit{Roofs}'. For each of these 500 textures, we take the visual texture and its associated heightfield as ground-truth pairs. 
% \faraz{\hl{Why use these 5 classes of textures?}}

For the stylization framework, we use Style2Fab~\cite{faruqi2023style2fab}, which allows users to personalize 3D models based on text prompts. We modified Style2Fab's system to take image prompts instead of text by changing the hyperparameters in the stylization module.

\subsection{Procedure}
For consistency, we perform stylization of a single tile of size 5$\times$5$\times$1 cm$^3$. To create the ground truth set of textures, we apply the heightfield from our dataset on the tile surface following the technique from Degraen et al.~\cite{degraen2021capturing}.  We take 50 random textures from our dataset (10\% of the dataset size) and stylize the tile with the texture image as the prompt. We subdivide the tile surface to 25k resolution for accurate texture generation and run the stylization process for 1500 iterations, as specified in Style2Fab~\cite{faruqi2023style2fab}. We apply stylization to only one face of the tile, the same as that of the ground truth textures, and freeze the geometry on the remaining faces, retaining a flat surface. This allows for a consistent comparison. Stylization iteratively modifies the geometry and color channel of the 3D model, and using the CLIP loss to assess the stylization quality. At the end of the study, we have 50 modified 3D tiles created using 50 random textures from our dataset. 

\subsection{Results:}
To quantitatively assess the fidelity of the stylized textures in replicating the ground-truth textures, we compare the Root Mean Square (RMS) values of the textures' heightfields as it has been shown to correlate to surface roughness~\cite{degraen2021capturing}. We take the 50 heightfields associated with the texture images used to stylize the 3D tiles. To extract the heightfield from the stylized tile, we take the boolean difference of original unstylized tile, and then map the displacement of the modified vertices onto the grayscale range(0 - 255). 
The RMS values capture the overall surface variation, allowing us to evaluate the differences between the original textures and the stylized outputs.

\begin{figure}[h]
    \centering
    \includegraphics[width=0.9\linewidth]{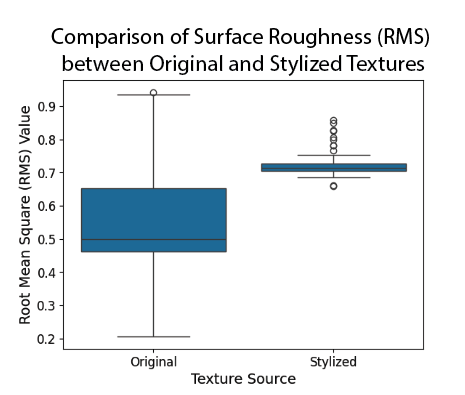}
    % \vspace{-3mm}
    \caption{The boxplot shows the distribution of Root Mean Square (RMS) values for original and stylized textures, representing surface roughness. The original textures exhibit a wider range of RMS values, indicating higher variability in surface roughness. In contrast, the stylized textures have consistently higher RMS values with less variability, indicating rougher and more uniform surfaces as a result of the stylization process.}
    % \vspace{-5mm}
    \label{fig:formative_study}
\end{figure}

Figure~\ref{fig:formative_study} presents a boxplot comparing the RMS distributions for the original textures and their stylized counterparts. We observe that the stylized textures generally exhibit higher RMS values compared to the original textures. RMS values can be interpreted as a metric for surface roughness~\cite{degraen2021capturing} suggesting that the stylization process results in rougher surfaces. Moreover, the RMS values for the original surface textures have a wider range of values, showing higher variability, whereas the stylized surfaces have lesser variability indicating more uniformity.  

To determine the statistical significance of the observed differences, we perform a Welch’s t-test between the RMS values of the original and stylized textures. The test reveals a statistically significant difference between the two groups ($t = 11.89$, $p < 0.0001$), indicating that the stylized textures have significantly different RMS values compared to the original heightfields.

This result suggests that Style2Fab and similar stylization strategies do not accurately modify the surface geometry to replicate a specific texture's heightfield. Further refinement in the stylization process could enhance the replication of texture variation for more accurate texture replication in 3D models. In the next sections, we present TactStyle, a system that allows creators to accurately replicate the tactile properties via a new geometry stylization approach. 
% \section{System Overview}

% Our system enables the generation of heightfields based on visual images of a texture, and integrates into an existing stylization method for 3D models. The core of our system is a modified image generation model, fine-tuned to generate heightfields from texture images. In this section, we describe TactStyle's user interface. 

\section{System Overview}

Prior work~\cite{faruqi2023style2fab, michel2022text2mesh, x_mesh} has shown that Generative-AI based stylization methods closely approximate the user's style visually. However, our formative study found that such geometry modifications do not accurately replicate the desired texture represented by its surface microgeometry. We designed TactStyle to enable the replication of a surface's microgeometry, and by extension, its tactile properties. 

% We designed TactStyle to focus on the geometry modification part of the stylization process, using Style2Fab~\cite{faruqi2023style2fab} as the color optimization method. 

\begin{figure*}[h]
    \centering
    \includegraphics[width=0.9\linewidth]{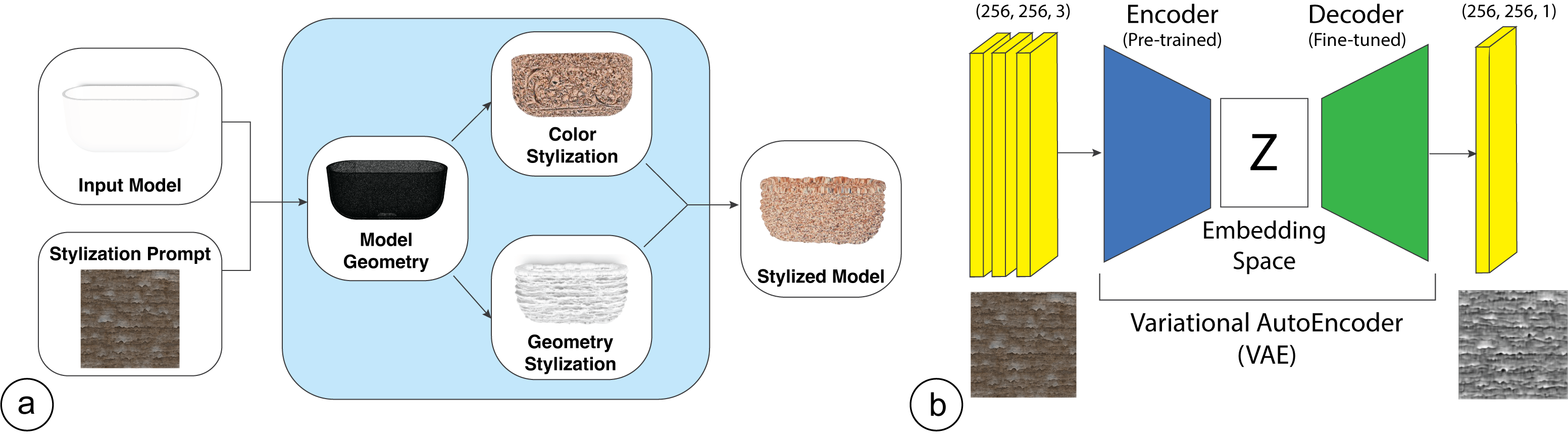}
    % \vspace{-3mm}
    \caption{TactStyle augments traditional 3D model stylization techniques by introducing a novel geometry stylization module that replicates the tactile properties of textures based on user input. (a) The system takes an input model and a stylization prompt (e.g., an image of a texture) and applies two separate stylization processes: (1) Color Stylization and (2) Geometry Stylization. The color stylization modifies the model's visual appearance, while the geometry stylization alters its surface to reflect tactile properties. The two modules operate in tandem, creating a stylized 3D model that replicates both the visual and tactile aspects of the texture. (b) The geometry stylization module uses a variational autoencoder (VAE) to generate heightfields from texture images, which are then applied to modify the model’s surface geometry, enabling co-optimization of geometry and color for a unified tactile and visual experience.}
    % \vspace{-5mm}
    \label{fig:system}
\end{figure*}

\changes{
TactStyle augments existing stylization methods with a new approach to modifying the geometry of 3D models that replicates the tactile properties of the texture described by the user. It accomplishes this by co-optimizing the geometry and the color channels separately. We call these two modules: (1) the color stylization module and (2) the geometry stylization module. The focus of this paper is on the geometry stylization module. 

Our main challenge was to design a geometry stylization module that modifies a 3D model's geometry to replicate the tactile properties of a texture. We leverage the fact that heightfields can be represented as images, and thus TactStyle accomplishes this goal by fine-tuning an image generation model to generate heightfields based on visual images of a texture. This heightfield is then used to modify the 3D model's geometry using an approach based on UV mapping. Thus, the color and geometry stylization modules work in tandem, stylizing the color and geometry of the 3D model to replicate both the visual appearance and tactile properties of a given texture (Figure~\ref{fig:system}a). }

% Users can access TactStyle through a simple user-interface plugin for the open-source 3D design software tool Blender~\cite{blenderUI}. To stylize a model with TactStyle, the user: (1)~loads the model, and (2)~provides an image prompt for their desired style. 

\section{Heightfield Generation Technique}
In this section, we describe our novel heightfield generation model. This model takes a texture image as a prompt and generates the associated heightfield. For this purpose, we fine-tune a trained diffusion-based Image-to-Image model and integrate it into the TactStyle system. In the following subsections, we describe the modified architecture of the diffusion model and the dataset used to train and test the system. 

\subsection{Diffusion Model}
We approach this problem as an image generation task and use a modified version of the Stable Diffusion model~\cite{rombach2021highresolution}, a popular open-source image generation model. Specifically, we use an image-to-image generation model proposed in SDEdit~\cite{meng2021sdedit}. This deep-learning model uses a diffusion model to synthesize new realistic images. Given an input image along with a user prompt in the form of text or image, SDEdit first adds noise to the input, then subsequently denoises the resulting image to generate a modified image based on the user prompt. At the core of this diffusion-based generative model is a variational autoencoder~\cite{kingma2013auto} (VAE), which encodes images into a latent representation, and decodes that latent representation into an image. 

The VAE is trained to encode an image into a latent representation, a compact high-dimensional representation that can then be `decoded' using another network called a `Decoder' to generate another image. More details on the architecture and training approach are available in Meng et al.~\cite{meng2021sdedit} and Kingma et al.~\cite{kingma2013auto}. Our goal with this model was to generate a heightfield given an image of a texture. Since heightfields are traditionally represented as grayscale images, we (1) modify the VAE architecture to generate representative grayscale images, i.e., heightfields, (2) and fine-tune the trained model on our texture image-heightfield pairs. 

\subsection{Modified Model Architecture}
\label{sec:model_architecture}
We use a trained open-source Image-to-Image Generation model available through the Diffusers library~\cite{von-platen-etal-2022-diffusers}. As described above, this model's essential component is the VAE, which encodes an image into a latent representation and then decodes it into another image. This VAE is structured to generate images in 3 (RGB) channels. We modify the architecture's decoder module by adding 4 additional layers to learn heightfield features and modify the final layer to output single-channel grayscale images. This approach was motivated by the fact that the pre-trained model was trained to generate colored images, and there are additional features that the model would need to learn to generate heightfield-specific features. 

In fine-tuning our modified image generation model, our goal was both to maximize the similarity in intensity between the target and generated heightfield and minimize their perceptual difference. For comparing overall intensities, we use the Mean Squared Error Loss (MSE), a standard in regression and image generation tasks. For the perceptual similarity metric, we use the Structural Similarity Index Measure (SSIM)~\cite{wang2004image}. These two loss functions serve two different purposes. MSE calculates an average of per-pixel similarity that provides a guide towards a similar intensity in generated images. However, independent training with MSE does not generate high-quality heightfields because it assumes pixel-wise independence. For instance, blurred images can have a large perceptual difference but a small MSE loss. SSIM on the other hand, takes into account the luminance, contrast, and structure of the two images being compared, highlighting local structural differences. Thus, a combination of these two loss functions allows us to generate heightfields that are similar in both overall intensity (MSE) and local structural features (SSIM). In training our model, we use these both loss measures.

\subsection{Training Methodology}

The Variational Auto Encoder (VAE) is fine-tuned for generating accurate heightfields using the PBR Dataset consisting of texture image-heightfield pairs. The associated heightfields serve as ground truth representations of the tactile features, and our model learns the correlation between visual appearance and tactile properties.

We fine-tune the model updating the decoder parameters over 60 epochs, using a batch size of 10 images with an RMSprop optimizer. We use a lower learning rate ($1e^{-5}$) for fine-tuning existing layers in the VAE model, and a higher learning rate ($1e^{-3}$) for the newer layers. This was done because the original layers are already trained and need small adjustments, whereas the newer layers are randomly initialized and require larger changes. Since our goal was to modify the `Decoder' module of the VAE, we froze the weights for the encoder module, training the weights for only the decoder module. 

\subsection{Dataset}
% \faraz{\hl{why these 5 classes of datasets?}}
To train our model on realistic textures, we utilize the CGAxis repository~\cite{cgaxis_pbr_20_parquets}, which contains a wide range of textures designed to provide accurate real-world simulations of materials in 3D environments. We collected 500 pairs of texture images and corresponding heightfields in 4k resolution. The dataset contains 5 different material types: `\textit{Parquets}', `\textit{Wood}', `\textit{Rocks}', `\textit{Walls}', and `\textit{Roofs}', containing 100 textures each. This allows for a diverse set of textures to train our model. For each of these 500 textures, we collect the visual texture and its associated heightfield as ground-truth pairs.  These heightfields represent the tactile features of the textures and are critical for learning the correlation between visual appearance and haptic properties. 

We split our dataset into a train and test set, using a 90\% - 10\% split, resulting in 450 textures to train our model and 50 textures to test it. We also augment the train set by rotating each image-heightfield pair by 90 degrees three times, effectively generating four variations for each texture and resulting in a total of 1,800 textures in our train set. This augmentation allows us to increase the diversity of the data, providing a more comprehensive set of examples for training our model without introducing synthetic artifacts. This enables the model to learn more robust and invariant representations of visual and tactile features, improving its ability to generalize across different orientations of textures.

\subsection{Texture Application}
\label{sec:texture_application}
To apply our textures to 3D models, we apply the heightmap by displacing vertices along their normals based on the corresponding height values from a UV map normalized to fit the texture map, producing a texturized object ready for 3D printing.
This process creates a final texturized object that is ready to be 3D printed, as shown in geometry stylization step of Figure~\ref{fig:system}a.

% %LOSS FUNCTION EQUATIONS
% % SSIM
% \[
% \text{SSIM}(x, \hat{x}) = \frac{(2 \mu_x \mu_{\hat{x}} + C_1)(2 \sigma_{x\hat{x}} + C_2)}{(\mu_x^2 + \mu_{\hat{x}}^2 + C_1)(\sigma_x^2 + \sigma_{\hat{x}}^2 + C_2)}
% \]

% where \(x\) and \(\hat{x}\) are the original and reconstructed images, \(\mu_x\) and \(\mu_{\hat{x}}\) are the means, \(\sigma_x^2\) and \(\sigma_{\hat{x}}^2\) are the variances, \(\sigma_{x\hat{x}}\) is the covariance, and \(C_1\) and \(C_2\) are constants to avoid division by zero.

% % MAE
% \[
% \mathcal{L}_{\text{MAE}}(x, \hat{x}) = \frac{1}{N} \sum_{i=1}^{N} |x_i - \hat{x}_i|
% \]

% where \(x_i\) and \(\hat{x}_i\) are the ground truth and predicted values, respectively, and \(N\) is the total number of data points.

% % BCE
% \[
% \mathcal{L}_{\text{BCE}}(x, \hat{x}) = -\frac{1}{N} \sum_{i=1}^{N} \left[ x_i \log(\hat{x}_i) + (1 - x_i) \log(1 - \hat{x}_i) \right]
% \]

% where \(x_i\) represents the true label, \(\hat{x}_i\) represents the predicted probability, and \(N\) is the number of samples.

\faraz{
\section{User Interface and Workflow}

\begin{figure}[h]
    \centering
    \includegraphics[width=0.9\linewidth]{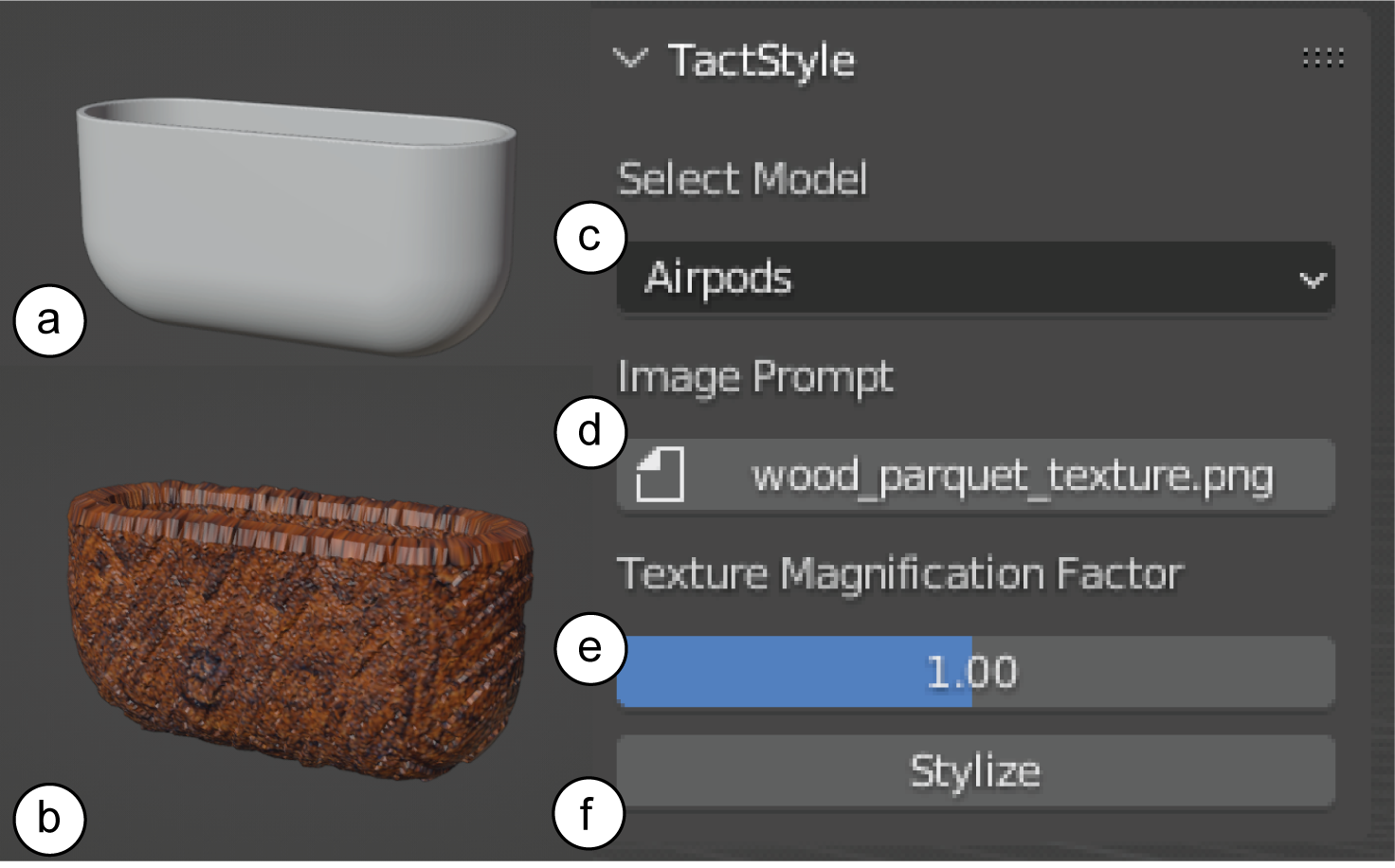}
    % \vspace{-3mm}
    \caption{TactStyle's user interface, implemented as a Blender plugin, allows users to load (a)~original 3D model and (b)~stylize with image prompts. In order to use TactStyle, the user (c)~loads the model, (d)~uploads an image of their desired texture,  (e)~optionally adjust the Texture Magnification Factor to control the level of height displacement applied on the 3D model. (f)~Finally, the user clicks the ``Stylize'' button, which starts the stylization process using TactStyle’s integrated color and geometry stylization modules.}
    % \vspace{-5mm}
    \label{fig:user_interface}
\end{figure}

TactStyle has been implemented as a plugin for the open-source 3D design software tool Blender~\cite{blenderUI} to allow easy integration with makers' existing workflows. Figure~\ref{fig:user_interface} shows a view of the interface. To stylize a model with TactStyle, the user (1)~loads their model, (2)~uploads the image prompt of the desired texture, and (3)~clicks the stylize button. TactStyle then processes the model and stylizes it using the integrated color and geometry simulation modules. The stylized model is rendered next to the original model, which the user can export for fabrication. 

\subsubsection{Preprocessing}
Once the user has loaded an OBJ file of their 3D mesh into the plugin, the model is automatically pre-processed for stylization. The model is first standardized to a unit-sized cube for stylization. Next, we use Pymeshlab~\cite{pymeshlab} to increase the model's resolution by subdivision to 25k faces following the standardization protocol from Style2Fab~\cite{faruqi2023style2fab}. This enables accurate stylization of the model by increasing the number of vertices on the model, which are then modified both in color and geometry to approximate the style desired by the user. 

\subsubsection{Stylization}
TactStyle used two modules — the color stylization module and the geometry stylization module to optimize both visual and tactile properties. As shown in Figure~\ref{fig:system}, TactStyle uses Style2Fab~\cite{faruqi2023style2fab} for iterative color optimization. Here the model's geometry is frozen, and the generative AI model modifies the color channels of the vertices to approximate the style in the image. Next, the geometry stylization module uses the modified image generation model to generate a heightfield using the texture image prompt provided by the user. This heighfield is applied on the model using the technique described in section ~\ref{sec:texture_application}. The completed model is rendered alongside the original model for review. Furthermore, the segmentation tool from Style2Fab~\cite{faruqi2023style2fab} has been integrated into TactStyle. This allows the user to have multiple textures on the same model. For this, the user can segment the model through Style2Fab's segmentation, and then apply TactStyle on individual segments. 

\subsubsection{Fine-Tuning and Export} Users can iterate on this process and apply new styles using new image prompts as needed. Users can also optionally increase the amount of height displacement to magnify their texture by changing the `Texture Magnification Factor' slider shown in Figure~\ref{fig:user_interface}e which can exaggerate or diminish the texture applied. This factor is by default at 1.0, which corresponds to the value used in our study following the height displacement values from Degraen et al.~\cite{degraen2021capturing}. 

Finally, the user can export the stylized model and fabricate it. }

% Mention that we didn't use this optional factor in our study, and just use the default values. 

\section{Technical Evaluation}
To validate the effectiveness of our system, we conducted a quantitative evaluation. We evaluate TactStyle's performance on its ability to replicate the surface micro-geometry, represented by the ground truth heightfield. We evaluate TactStyle's results using two metrics: (1) the RMS Error, which represents the difference in surface roughness, (2) the Mean Squared Error (MSE) which calculates the average error in per-pixel intensity between the textures. In the following subsections, we discuss the results of the quantitative evaluation.

\begin{figure*}[h]
    \centering
    \includegraphics[width=0.6\linewidth]{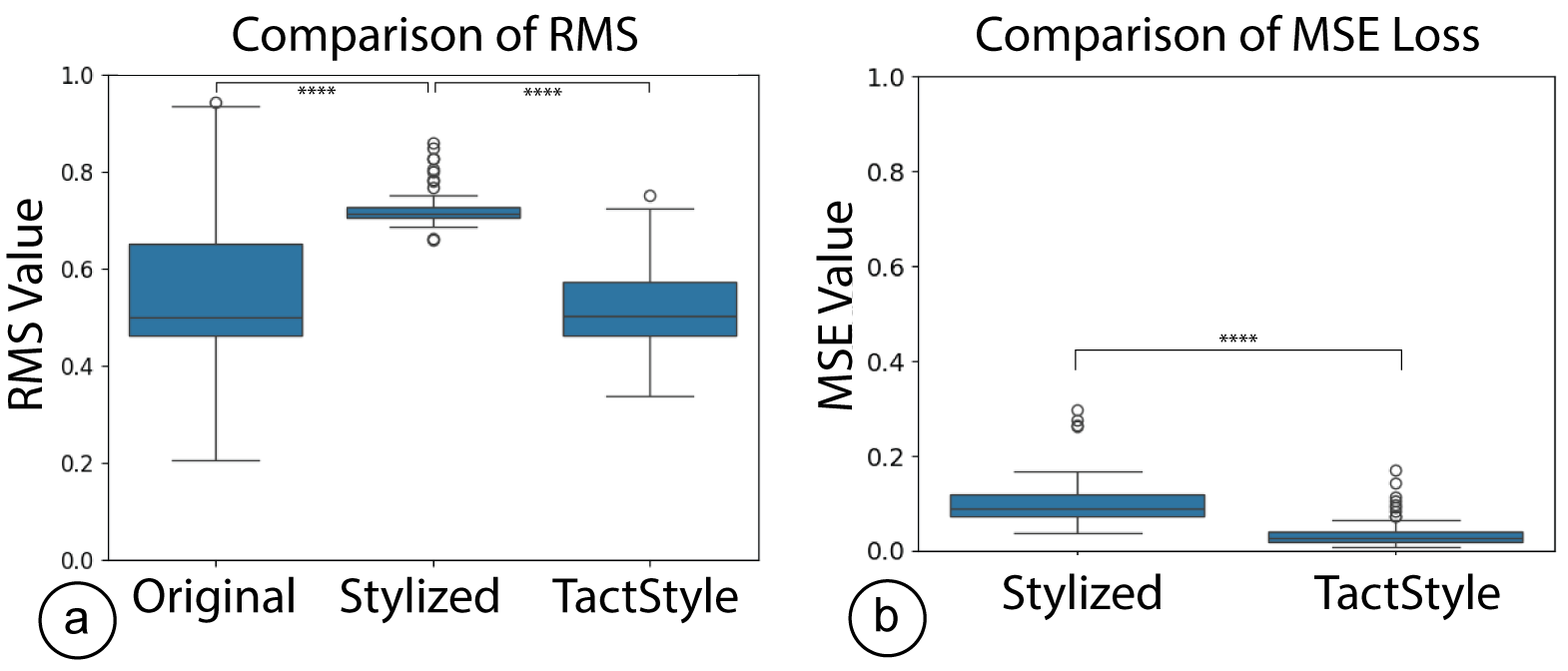}
    % \vspace{-3mm}
    \caption{Quantitative Comparison of TactStyle and Stylized Textures, and original heightfields. (a) Comparison of RMS values for Original, Stylized, and TactStyle textures, demonstrating that TactStyle replicates surface roughness more closely to the Original textures. (b) Box plot of MSE loss between the Original textures and the Stylized and TactStyle textures. TactStyle exhibits significantly lower MSE compared to the stylized method, indicating more accurate texture replication.  (****p < 0.0001)
    }
    % \vspace{-5mm}
    \label{fig:tech_eval}
\end{figure*}

\subsubsection{Analyzing Root Mean Square Error:}
We evaluate the Root Mean Square (RMS) values of the generated heightfields, which allow us to compare the overall surface roughness of the generated textures~\cite{degraen2021capturing}. Figure~\ref{fig:tech_eval}a shows the comparison between the Original, Stylized, and TactStyle textures. The original textures exhibit a wide range of RMS values, reflecting the inherent variability in surface roughness across different textures. 

To evaluate the differences in RMS values between the Original, Stylized, and TactStyle textures, we performed a Welch's ANOVA test which indicated a statistically significant  difference ($F = 47.58$, $p < 0.0001$). Next we conducted a Games-Howell post-hoc analysis. We found a significant difference between Stylized and Original textures ($T = 6.79$, $p < 0.0001$) and TactStyle and Stylized ($T = 14.34$, $p < 0.0001$) textures. However, we found no significant difference between the TactStyle and Original textures ($p > 0.05$). These results suggest that stylization process results in textures with significantly higher surface roughness compared to the original and TactStyle's generated textures.

\subsubsection{Analyzing Mean Squared Error:}
 MSE measures per-pixel intensity differences, where 0 indicates identical per-pixel intensities, and 1 indicates completely different intensities. As shown in Figure~\ref{fig:tech_eval}b, TactStyle exhibits lower MSE (M = 0.03, std-dev = 0.03) compared to the stylized method (M = 0.10$, $std-dev = 0.05), indicating more accurate texture replication. To evaluate statistical significance, we conducted a Welch’s t-test, and found that the MSE Loss for TactStyle's results was significantly lower than that of Stylized results ($F = 6.79$, $p < 0.0001$).

 % \faraz{\hl{add a conclusion to this section}}

 % \subsubsection{Analyzing Structural Similarity Measure:} As discussed in ~\autoref{sec:model_architecture} SSIM is a perceptual metric used to measure the similarity between two images by comparing their structural information. It evaluates how similar the two images are in terms of luminance, contrast, and structure. A value of 1 for SSIM signifies a perfect replica in structure, while a value of 0, a totally different structure. We compared the Stylized and TactStyle's generated heightfield with the original heightfield.  As shown in Figure~\ref{fig:tech_eval}c, TactStyle (M = 0.42, std-dev=0.02) performs better than Stylized (M = 0.02, std-dev=0.002) at approximating the structural information in the original heightfield. To evaluate statistical significance, we conducted a Welch’s t-test, and found that the SSIM was significantly better for TactStyle's results than Stylized's results ($F = 107.76$, $p < 0.0001$). 
\section{Perception Study}
In order to evaluate TactStyle's accuracy at replicating the tactile feedback of textures, we performed a psycho-physical experiment used to evaluate texture replication techniques~\cite{degraen2021capturing, perez2016optimization}. The goal of our study was to understand if the reconstructed heightfield from TactStyle creates similar tactile perceptions to the original heightfield. In addition, our second goal was to evaluate if the tactile perception of TactStyle's heightfields are similar to the expected tactile perception from just looking at a visual image of the texture. To compare tactile properties, we take a representative set of descriptors from Degraen et al.~\cite{degraen2021capturing}.

% TactStyle is a system that takes a texture image and generates a heightfield that can be applied to a 3D model during stylization. This heightfield generation technique assumes a correlation between the visual properties of a texture, and their tactile properties, which TactStyle approximates using the fine-tuned image-generation model. 

\begin{figure*}[h]
    \centering
    \includegraphics[width=0.9\linewidth]{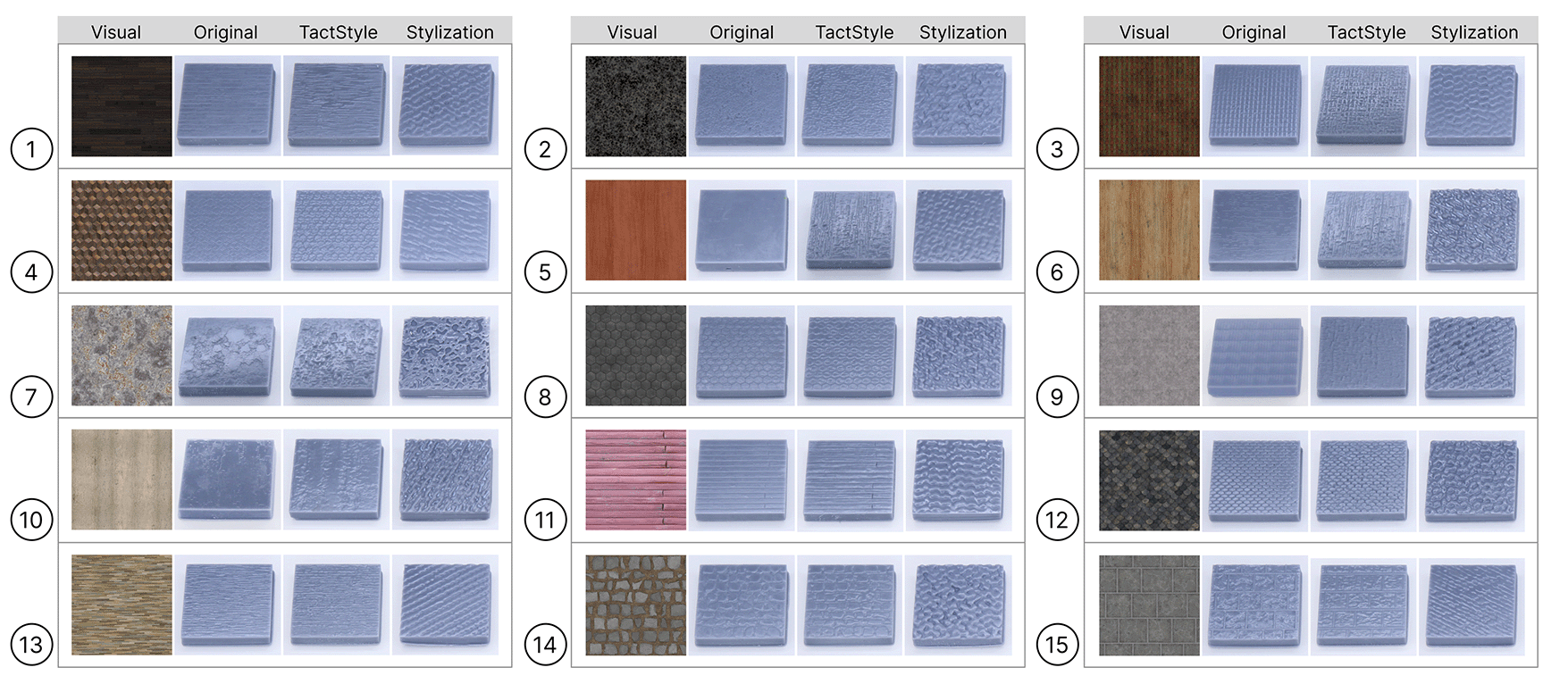}
    % \vspace{-3mm}
    \caption{3D-Printed samples of 15 textures from our test set used in perception study: We created four different sets for our perception study: the `visual set', `original set', `TactStyle set', and the `stylization set'. The original set was created with the heightfield associated with the texture and served as the groundtruth. The reconstructed set was created using TactStyle, with the texture image as input. The visual set was created using printed texture images. Finally, the stylization (baseline) set was created using Style2Fab, using the texture image as input. }
    % \vspace{-5mm}
    \label{fig:study_models}
\end{figure*}

\subsection{Conditions}
\label{sec:conditions}
\faraz{We created four distinct conditions to evaluate the tactile and visual characteristics of textures. These conditions were designed to isolate specific aspects of perception, allowing us to better understand how each modality contributes to the overall experience of texture replication. The conditions were:}

% \faraz{\hl{give a rationale behind analyzing visual and tactile characteristics separately}}

\begin{enumerate}
    \item \textbf{Visual (No Heightfield):} The texture image was printed with no heightfield on glossy paper and pasted on flat tiles. 
    \item \textbf{Original Heightfield:} The texture was printed with the heightfield originally provided with the texture (groundtruth).
    \item \textbf{TactStyle:} The texture was printed with the heightfield created from TactStyle with the texture image as input.
    \item \textbf{Stylization:} The texture was printed with the heightfield generated from Style2Fab~\cite{faruqi2023style2fab} using the texture image as input.
\end{enumerate}

\faraz{This separation of conditions was motivated by the literature on visuo-haptic stimuli integration by humans. When humans explore objects with their hands, vision and touch both provide information for estimating the properties of the object~\cite{ernst2002humans}. Vision frequently dominates the integrated visual–haptic percept~\cite{fleming2014visual, fleming2017material}. To address this, we kept the visual and tactile conditions separate to mitigate cross-modality influence and isolate modality-specific effects to assess tactile perception. Here, condition 4 (stylization) is our baseline to be compared with TactStyle's results. }

In this experiment, we investigate the following research questions:  

% \faraz{\hl{change the RQs}}

\faraz{
\begin{enumerate}{
    % \item \textbf{RQ1:} Does stylization replicate tactile properties of the original texture, represented by its heightfield?
    % \item \textbf{RQ2}: Does stylization replicate expected tactile properties of the original texture, represented by a texture's visual image?
    \item \textbf{RQ1:} How accurately does TactStyle replicate the tactile properties of a texture, as represented by its original heightfield?
    \item \textbf{RQ2:} To what extent do TactStyle-generated textures align with user expectations based on their visual appearance?
    \item \textbf{RQ3:} How do the tactile expectations derived from a texture’s visual appearance differ from its actual tactile properties, as represented by the heightfield?
    }
\end{enumerate}
}

\subsection{Dataset}
We collected 15 random samples from our test set, matching the size of the sample set used by Degraen et al.~\cite{degraen2021capturing} in their study on the perceptual similarity between real textures and their digital replicas. As these textures were not used to train the model, they can be used to evaluate TactStyle's ability to replicate unseen textures. Figure~\ref{fig:study_models} shows the models used in our study. This gave us a total set of 60 models (4 conditions for each of the 15 textures). All the conditions were presented to the user on tiles of the same size - 5cm x 5cm x 1cm. Our models were printed using an SLA printer, namely, Elegoo Saturn 3 Ultra. In order to keep our printed objects comparable to the previous studies, we used the Elegoo Resin Standard 2.0 – Grey, which has a Shore Hardness of 80-86 (Scale D). For comparison, Degraen et al.~\cite{degraen2021capturing} used a material with Shore Hardness of 83-86 (Scale D). We printed all the samples at a layer resolution of 30 $\mu$m.

\subsection{Study Design}
% We collected 15 random samples from our test set, matching the size of the sample set used by Degrean et al.~\cite{degraen2021capturing} in their study. 
We used a within-subjects experimental design. To control for carry-over effects, we counter-balanced conditions using round-robin ordering between participants. Our study was structured as a self-assessment test in which participants compared and recorded perceptual attributes of the 3D-printed texture samples from the different conditions.

% \faraz{\hl{Describe why these metrics were used for the perception study}}
During the study, each participant recorded their ratings of the sample in terms of hardness, roughness, bumpiness, stickiness, scratchiness, uniformity, and how isotropic the surface is, each on a 1-to-9 Likert scale, \faraz{1 indicating a low assessment and 9 indicating a high assessment of the respective variable.
To rate these dimensions, participants were asked the following questions:}

% \faraz{Hardness, Roughness, Bumpiness, Stickiness, and Stickiness are motivated from the Nagano et al.\cite{okamoto2012psychophysical}'s review on tactile dimensions of materials. Bumpiness refers to macro-roughness, and scratchiness is micro-roughness. By including these, }

% A value of 1 indicated low agreement, and 9 indicated high agreement. We exclude the descriptor `hairiness', since none of the textures in our dataset were representative of it.

\begin{enumerate}
    \item[Q1:] How \textbf{hard} does this surface feel? (1 meaning extremely soft, 9 meaning extremely hard)

    \item[Q2:] How \textbf{rough} does this surface feel? (1 meaning extremely smooth, 9 meaning extremely rough)

    \item[Q3:] How \textbf{bumpy} does this surface feel? (1 meaning extremely \faraz{flat}, 9 meaning extremely bumpy)

    \item[Q4:] How \textbf{sticky} does this surface feel? (1 meaning extremely slippery, 9 meaning extremely sticky)

    \item[Q5:] How \textbf{scratchy} does this surface feel? (1 meaning extremely dull, 9 meaning extremely scratchy)

    \item[Q6:] How \textbf{uniform} does this surface feel? (1 meaning extremely irregular, 9 meaning extremely uniform)

    \item[Q7:] How \textbf{isotropic} does this surface feel? (1 meaning extremely anisotropic, 9 meaning extremely isotropic)
    
    % \item[Q8:] What kind of material is this? (Open question)
\end{enumerate}

\faraz{Hardness, Roughness and Stickiness are motivated by related work indicating these are the base dimensions of tactile discrimination~\cite{hollins2000individual, okamoto2012psychophysical, yoshioka2007perception, vardar2019fingertipmetrics, hollins1993perceptual}.
Bumpiness and Scratchiness are informed by the notion that roughness can be divided into respectively macro and micro dimensions~\cite{okamoto2012psychophysical}.
The inclusion of Uniformity and Isotropy stems from the fact that our textures embed some directionality and localized variations, which affect perception during tactile exploration~\cite{degraen2021capturing}.
While other works have considered the additional dimension of Hairiness~\cite{degraen2021capturing, degraen2019enhancing}, we excluded this descriptor since none of our textures in our dataset were representative of it.}

\subsection{Apparatus}
\begin{figure}[h]
    \centering
    \includegraphics[width=0.9\linewidth]{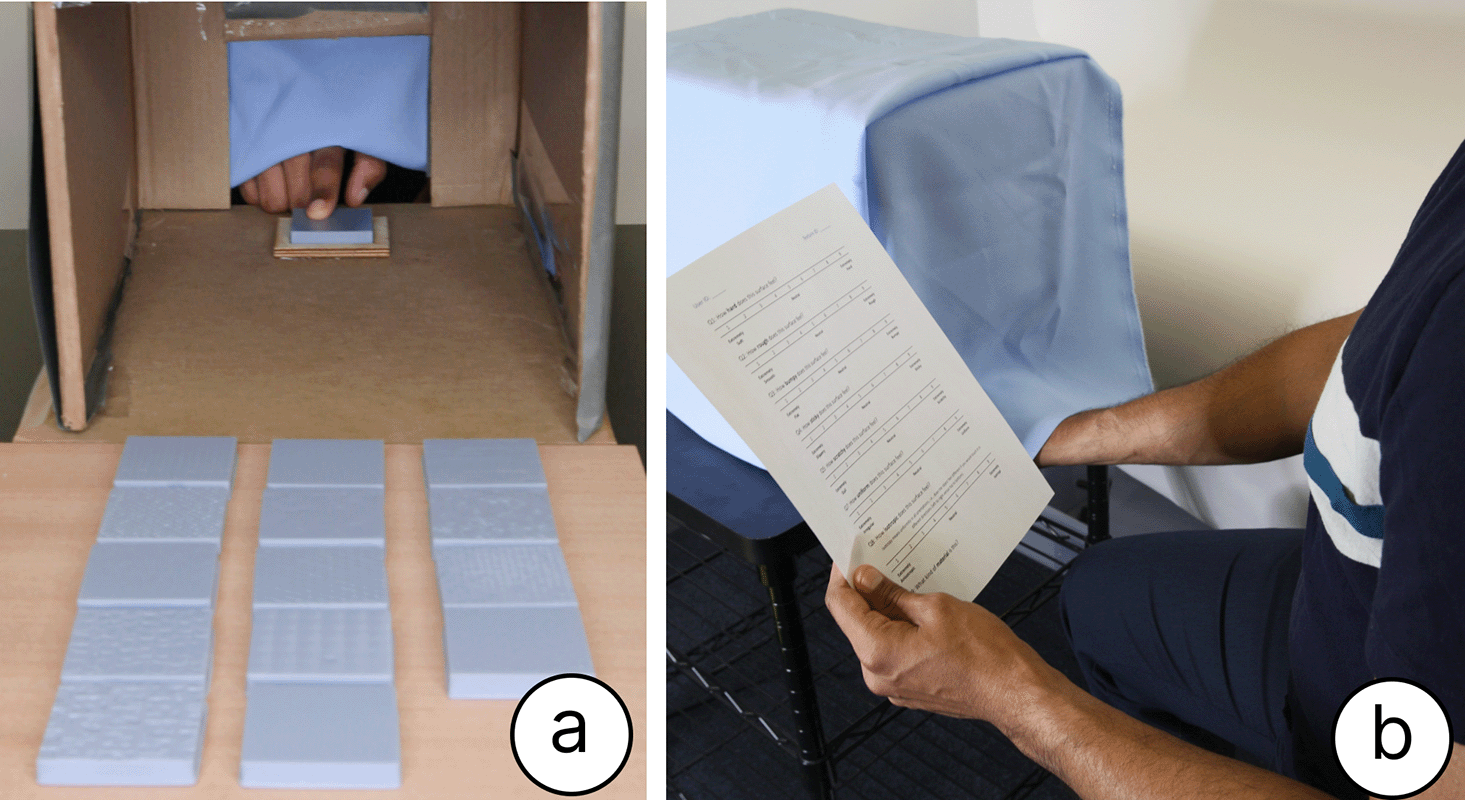}
    % \vspace{-3mm}
    \caption{Experimental Setup for perception study: a) Experimenter side, b) Participant side}
    % \vspace{-5mm}
    \label{fig:study_setup}
\end{figure}

Our apparatus was built to limit visual cues and ensure accurate recording of purely tactile perceptual attributes of the textures. Participants were positioned in front of a screen that separated them from the experimenter, as shown in Figure~\ref{fig:study_setup}. A small opening in the screen, covered by a piece of cloth, allowed participants to reach through and access the samples, placed by the experimenter. On the other side, the experimenter arranged the samples for the participants to explore. The samples were held in place with a laser-cut wooden frame. 

\subsection{Participants}
A total of 15 participants (6 female, 9 male, 22 - 38 years, $M=27.3$ years, $SD=5.4$) were recruited for our study. When asked about their hand dominance, 14 participants indicated to be right-handed with 1 participant indicating ambidexterity. All participants chose to use their right-hand index finger for the study. They were informed that they could only use this finger throughout the study \faraz{for consistency in perception}. All participants indicated that they do not suffer from any impairment to haptic perception to best of their knowledge. Participants were compensated with \$20 an hour for the 90-minutes long study.
% When asked about how often they performed

\subsection{Study Procedure}
The total study duration was set to be 90 minutes. At the start, participants were asked to fill in a consent form, and a short survey about their demographic data. Next, to ensure that participants clearly understood the perceptual descriptors used in the study, we conducted a short training session where participants were allowed to explore exemplar textures separate from the set being investigated. Once they were confident in their understanding of the descriptors, we proceeded to the evaluation stage. 

%We use the same descriptors. \faraz{cite the section where you introduced this concept.} 
 
During this stage, the experimenter placed one sample at a time at a fixed location behind a screen. The participant could insert their hand into the screen and feel the texture but were not allowed to see the texture itself. The participant was then asked to explore the texture and rate its tactile properties based on the 7 descriptors. During the visual perception stage, the samples were placed on a board next to the screen where the participants could see the texture but were not allowed to touch them. \faraz{This allowed us to isolate visual and tactile perception for textures evaluated in the study, and mitigate cross-modal influence~\cite{fleming2014visual, fleming2017material}. Prior work has shown that human fingers are particularly sensitive in perceiving and distinguishing textures~\cite{skedung2013feeling}. Since all the participants chose their right hand for the study, all participants were requested to use only their right index finger throughout their study for consistency.} The interaction window was limited to 5 seconds per sample so that the participants' first impressions could be communicated. All participants answered the descriptor questions for all 60 textures. 

Ethical approval for this study was obtained from the Ethical Review Board of the author's institute.

 % \subsection{Results: Touching}

%  \begin{figure*}[h]
%     \centering
%     \includegraphics[width=0.9\linewidth]{Figures/Correlations.png}
%     % \vspace{-3mm}
%     \caption{Box Plots showing the individual assessments on Hardness, Roughness, Bumpiness, Stickiness, Scratchiness, uniformity, and Isotropy}
%     % \vspace{-5mm}
%     \label{fig:study_models}
% \end{figure*}

\section{Results}

 \begin{figure*}[h]
    \centering
    \includegraphics[width=\linewidth]{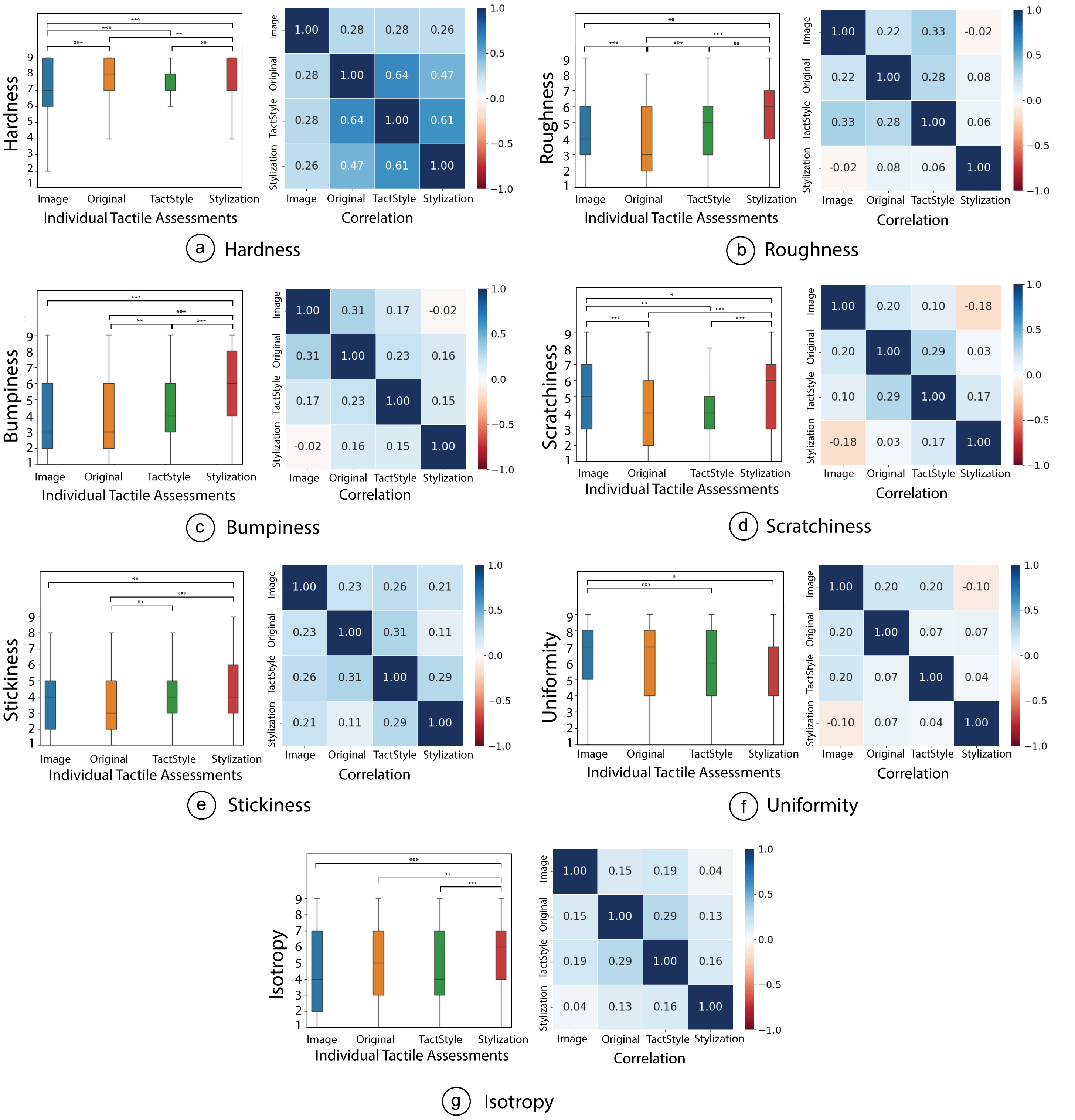}
    % \vspace{-3mm}
    \caption{\changes{Box Plots showing the individual assessments on Hardness, Roughness, Bumpiness, Stickiness, Scratchiness, Uniformity, and Isotropy. Tactile Correlations are shown as heatmaps showing the correlations between the 4 conditions for each descriptor.  (*p < 0.05, **p < 0.01, ***p < 0.001) }}
    % \vspace{-5mm}
    \label{fig:box_plots_correlations}
\end{figure*}

In the following section, we describe the analysis and the obtained results from our texture perception study. 

% \faraz{\hl{check numbers; add discussion in each subsection; shorten by following same format. }}

\changes{
\subsection{Comparing Visual and Tactile Ratings}
In this section, we present the results from the perception study. To analyze the individual tactile assessments, we conducted Friedman tests with post-hoc analysis using Wilcoxon signed-ranks tests and Bonferroni-Holm correction. Figure~\ref{fig:box_plots_correlations} shows box plots for each assessment. The assessment data for each descriptor is provided in Appendix~\ref{sec:appendix_perception_diff}. 

\subsubsection{Hardness}
\label{sec:diff_hardness}

Users perceived significant differences in Hardness between Original vs. Visual, Original vs. Stylization, Visual vs. TactStyle, Visual vs. Stylization, and TactStyle vs. Stylization. However, users did not perceive significant differences in Original vs. TactStyle. 

% I'm a bit confused here. I think you are repeating numbers in the second part, right? I suggest to keep the first sentences with the numbers, and the last sentences more 'human'-readable, for example: "This indicates that Stylization created harder samples than..." Also, the averages are not indicated, this may be important to indicate direction.

\subsubsection{Roughness}
\label{sec:diff_roughness}

Users perceived significant differences in Roughness in all comparisons except between the Visual and the TactStyle conditions.

% Thus, no significant difference was found between Visual vs. TactStyle ($p > 0.05$), but significant differences were observed between Original vs. TactStyle, Original vs. Stylization, and TactStyle vs. Stylization. 

\subsubsection{Bumpiness}
\label{sec:diff_bumpiness}

Users perceived no significant differences in bumpiness between Original vs. Visual, or Visual vs. TactStyle. However, they perceived significant differences between Original vs. TactStyle, Original vs. Stylization, Visual vs. Stylization, and TactStyle vs. Stylization.

\subsubsection{Scratchiness}
\label{sec:diff_scratchiness}

Users perceived significant differences in Scratchiness between Original vs. Visual, Original vs. Stylization, Visual vs. TactStyle, Visual vs. Stylization, and TactStyle vs. Stylization. However, they did not perceive any significant difference between Original vs. TactStyle.

\subsubsection{Stickiness}
\label{sec:diff_stickiness}

Users perceived no significant difference in Stickiness between Original vs. Visual, Visual vs. TactStyle, or TactStyle vs. Stylization conditions. However, significant differences were found between Original vs. TactStyle, Original vs. Stylization, and Visual vs. Stylization.

\subsubsection{Uniformity}
\label{sec:diff_uniformity}

Users perceived no significant differences in uniformity between Original vs. Visual, Original vs. Stylization, Original vs. TactStyle, or TactStyle vs. Stylization. However, significant differences were found between Visual vs. TactStyle and Visual vs. Stylization.

\subsubsection{Isotropy}
\label{sec:diff_isotropy}

Users perceived no significant difference in isotropy between Original vs. Visual, Original vs. TactStyle, or Visual vs. TactStyle samples. However, they perceived significant differences between Original vs. Stylization, Visual vs. Stylization, and TactStyle vs. Stylization.

% This shows that Stylization samples were perceived significantly different from Original and Visual samples, while TactStyle samples were not.

\subsection{Perceptual Correlations}
To uncover relationships between different tactile perceptions in our samples, we performed Spearman's rank-order correlation analysis. For each descriptor, we evaluate the relationship between different texture descriptors across the Original, TactStyle, Visual, and Stylization samples. This analysis helped determine whether the tactile ratings of the textures were correlated across different conditions. Figure~\ref{fig:box_plots_correlations} shows correlation plots for each assessment. The assessment data for each descriptor is provided in Appendix~\ref{sec:appendix_perception_corr}.  

% \faraz{\hl{check numbers; add discussion in each subsection; shorten by following same format}}

\subsubsection{Hardness}
\label{sec:corr_hardness}

There were significant correlations in perception of hardness between all pairs of textures. This can be explained by the fact that the samples were printed with the same material. However, as described in Section~\ref{sec:diff_hardness}, users perceived significant differences between all pairs except Original vs TactStyle samples. This suggests that TactStyle effectively replicates the tactile properties of hardness from the Original textures, while stylization does not.

\subsubsection{Roughness}
\label{sec:corr_roughness}
There were significant correlations in perception of roughness between Original vs Visual, Original v TactStyle, and Visual vs TactStyle. We found in Section ~\ref{sec:diff_roughness} that users perceived significant differences in roughness perception in all comparisons except Visual vs TactStyle condition. Thus, TactStyle effectively replicates tactile perception of roughness from visual expectations of a texture.

\subsubsection{Bumpiness}
\label{sec:corr_bumpiness}

There were significant correlations in perception of bumpiness between Original vs Visual, Original vs TactStyle and Visual vs TactStyle. However, we found significant differences in comparing Original vs TactStyle (Section \ref{sec:diff_bumpiness}), but not in Visual vs TactStyle, and Original vs Visual. Thus, TactStyle effectively replicates bumpiness of textures from their visual expectations.

\subsubsection{Scratchiness}
\label{sec:corr_scratchiness}
There were significant correlations in perception of scratchiness between Original and Visual, and Original and TactStyle. We found that users did not perceive significant differences between Original and TactStyle samples (Section \ref{sec:diff_scratchiness}). Thus, TactStyle effectively replicates Scratchiness of Original textures.

\subsubsection{Stickiness}
\label{sec:corr_stickiness}
There were significant correlations in perceived stickiness of textures between Original and Visual, Original and TactStyle, Visual and TactStyle, Visual and Stylization. Since users did not perceive any significant differences in perceived stickiness (Section~\ref{sec:diff_stickiness}) for Visual and TactStyle, this shows that TactStyle effectively replicates perceived stickiness of textures from their visual expectations.

\subsubsection{Uniformity}
\label{sec:corr_uniformity}
There were significant correlations in perceived uniformity, between Original and Visual, Original and TactStyle, and Visual and TactStyle conditions. Since we found that users did not perceive any significant differences in perceived uniformity between Original and TactStyle (Section~\ref{sec:diff_uniformity}), TactStyle effectively replicates perceived uniformity of textures from their Original textures.

\subsubsection{Isotropy}
\label{sec:corr_isotropy}
There were significant correlations in perceived isotropy between Original and TactStyle, Original and Visual, and Visual and TactStyle samples. Since users did not perceive any significant differences between Original vs TactStyle, Original vs Visual, and Visual vs TactStyle (Section~\ref{sec:diff_isotropy}), this shows us that TactStyle is able to effectively replicate perceived isotropy from both Visual expectations and Original textures.

\subsection{Discussion}
We conducted a comparative analysis of the 4 texture sets - Visual, Original, TactStyle, and Stylization, identifying which textures are significantly different on various tactile descriptors, and which of them are correlated. We found that TactStyle effectively replicates visual and Original textures for several key descriptors, and outperforms the baseline stylization method. Based on our analysis, we were able to answer all our research questions stipulated in Section~\ref{sec:conditions}.

% \faraz{\hl{change format to follow new RQs; add conclusion section on what these mean. }}

% \subsubsection{RQ1: Stylization does not accurately replicate Original Textures:}
% The analysis reveals that Stylization does not perform well in replicating the tactile features of Original textures. In between Stylization samples and Original samples, while Hardness shows a moderate correlation, significant differences across most descriptors, such as Roughness, Bumpiness, Stickiness, and Scratchiness, indicate poor alignment between the tactile experiences provided by Stylization and the Original textures.

% \subsubsection{RQ2: Stylization does not accurately replicate Visual Textures:}
% While showing correlation for stickiness, Stylization does not effectively replicate the tactile expectations from Visual samples, with most descriptors showing significant differences and weak or no correlations. 

\subsubsection{RQ1: TactStyle accurately replicates several Original Texture Descriptors:}
TactStyle effectively replicated the tactile experiences expected from visual cues for several descriptors. Hardness, Scratchiness, Uniformity, and Isotropy are correlated between Original and TactStyle samples without exhibiting significant perceptual differences. This suggests that TactStyle is capable of closely replicating the tactile sensations associated with these descriptors from the original textures. 

% Uniformity, however, does not show a significant correlation despite not having significant differences between conditions.

The analysis also showed that Stylization does not perform well in replicating the tactile features of Original textures. In between Stylization and Original samples, significant differences across most descriptors such as Hardness, Roughness, Bumpiness, Stickiness, and Scratchiness, indicate poor alignment between the tactile experiences replicated by Stylization compared to the Original textures.

\subsubsection{RQ2: TactStyle accurately replicates several Visual Texture Descriptors:}
TactStyle effectively replicates several tactile features based on the visual textures, as demonstrated by significant correlations and the absence of perceptual differences in descriptors such as Roughness, Bumpiness,  Stickiness, and Isotropy. This indicates that TactStyle can reproduce the tactile experiences expected from visual cues based on these descriptors. 

In contrast, stylization does not reliably replicate tactile expectations derived from visual samples. Stylization does not effectively replicate the tactile expectations from Visual samples, with most descriptors showing significant differences.

\subsubsection{RQ3: Differences Between Tactile Expectations from Visual Textures and Actual Tactile Perceptions:}
We also compared the expected tactile perceptions of visual textures, and the tactile perceptions of the original textures. We found that most descriptors such as Bumpiness, Stickiness, Uniformity, and Isotropy align closely between visual perceptions and actual original surface tactile experiences. In contrast, descriptors like Hardness, Roughness, and Scratchiness show significant perceptual differences between Visual and Original samples, although moderate correlations indicate some level of consistency. These results suggest that while certain tactile experiences can be anticipated based on visual cues alone, others may require direct tactile interaction for accurate perception. 

}

% For Hardness, a significant difference was observed between Visual and Original samples ($p < 0.001$), but a moderate positive correlation was found (r = 0.282802, $p < 0.001$), indicating some alignment between the expected and actual tactile experiences. In the case of Roughness, no significant difference was found between Visual and Original samples ($p > 0.05$), with a significant correlation observed (r = 0.215618, $p < 0.01$). For Bumpiness, no significant difference was found between Visual and Original samples ($p > 0.05$), but a strong correlation was detected (r = 0.312852, $p < 0.001$), suggesting a high level of agreement between the two conditions. Stickiness followed a similar trend, with no significant difference between Visual and Original samples ($p > 0.05$) and a significant correlation (r = 0.225262, $p < 0.01$). Scratchiness presented a significant difference between Visual and Original samples ($p < 0.001$), but still showed a moderate correlation (r = 0.195008, $p < 0.01$). Finally, for Isotropy, a significant correlation was observed between Visual and Original samples (r = 0.154919, $p < 0.05$), despite no significant difference between them ($p > 0.05$). Uniformity, however, showed no significant correlation or difference between Visual and Original samples ($p > 0.05$).

\section{Applications}

\begin{figure*}[h]
    \centering
    \includegraphics[width=0.9\linewidth]{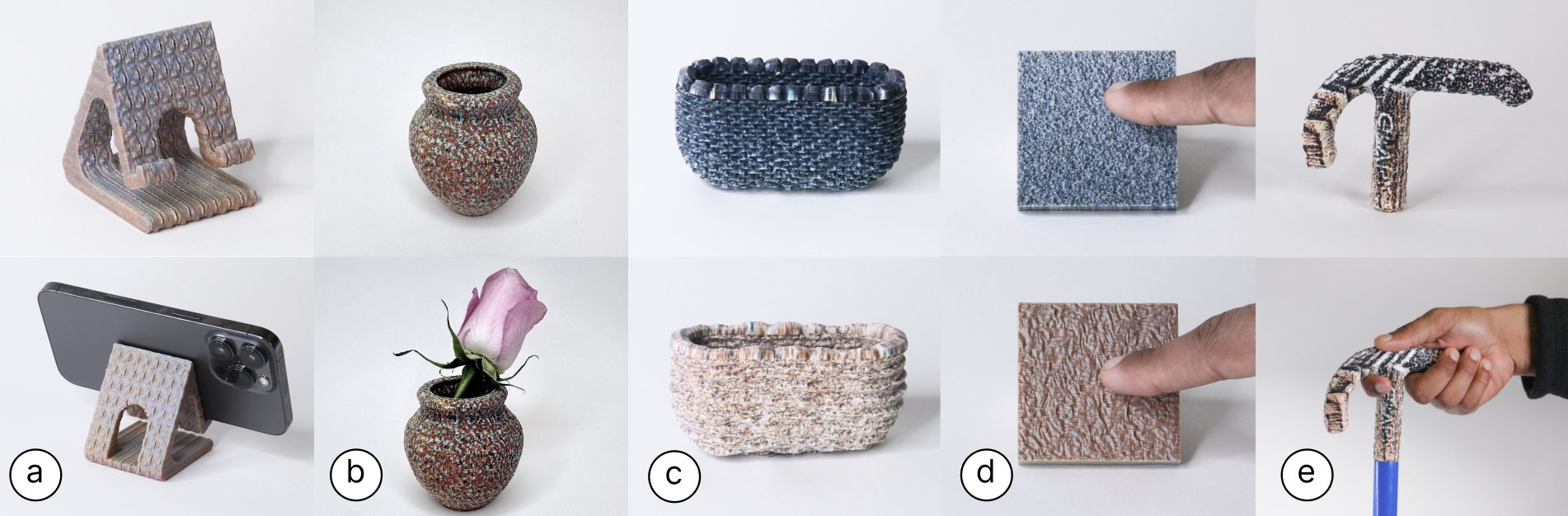}
    % \vspace{-3mm}
    \caption{Application Examples for TactStyle: a)~a phone stand stylized with a wood-parquet texture, b)~a granite-textured vase, c)~an airpods case stylized with a texture of `round stone roof' and `layered brown rock', d)~two tiles, one styled with a volcanic rock texture and the other styled with stone from the Grand Canyon, e)~walking stick handle stylized with a rough rock texture.}
    % \vspace{-5mm}
    \label{fig:application_examples}
\end{figure*}

In this section, we showcase how TactStyle's stylization technique allows users to stylize 3D models with accurate tactile properties for fabrication. We demonstrate five application scenarios across four categories: home decor, personal accessories, tactile learning tools, and personalized health applications. All 3D models shown in Figure~\ref{fig:application_examples} were stylized with TactStyle \faraz{using textures not present in the training dataset} and printed on a Stratasys J55 printer.  

\subsection{Home Decor}

TactStyle can be used to apply textures to objects downloaded from platforms like Thingiverse, enabling users to enhance the tactile experience of 3D-printed items at home. This allows individuals to create customized, textured versions of everyday objects, adding both aesthetics and functionality. We illustrate two applications of TactStyle in applying textures to functional home objects. Figure~\ref{fig:application_examples}a we shows a wood-parquet textured phone stand, demonstrating how organic textures can be applied to enhance the visual appeal and usability of frequently handled items. Figure~\ref{fig:application_examples}b, shows a granite-textured vase. By combining TactStyle and digital fabrication techniques, users can now personalize their objects, or prototype specific tactile properties in addition to the aesthetics of their home decor objects.  

\subsection{Personalizing Accessories}
Personal accessories are a popular domain for personalized fabrication. TactStyle enables creators to replicate both the `look' and the `feel' of textures based on image input, allowing creators to create customized versions of their accessories with specific textures and fabricate them with digital fabrication. \faraz{In Figure~\ref{fig:application_examples}c we showcase an AirPods case, stylized with two different textures: a round stone roof texture taken from an image of a `round stone roof' (top), and another stylized with an image of a `layered brown rock'. These textures not only provide visual distinction but also have different surface microgeometry, associated with the texture}.

 \subsection{Tactile Learning Tools}
TactStyle has the potential to create educational tools that enhance learning in subjects such as geometry, topography (e.g., the texture of different terrains), and biology (e.g., the texture of animal skins). To exemplify this concept, we present two examples in Figure~\ref{fig:application_examples}d: the top surface features a `volcanic rock texture', and the bottom surface replicates the texture of stone from `the Grand Canyon'. \faraz{Both textures are not present in the dataset, however are samples of a class that TactStyle is trained on. Thus, TactStyle is able to effectively generalize over different texture classes provided they were represented in the training data.} Tangible learning materials are well-known to improve educational outcomes, particularly by engaging multiple senses \cite{gonzalez2019tangible}. TactStyle offers a new way to create such materials, allowing educators to bring textures and surfaces to life in the classroom. By giving students the ability to physically interact with these textures, TactStyle could potentially help them better understand the tactile properties of objects, making abstract concepts more concrete and accessible.

\subsection{Customizable Assistive Devices}
In the field of ``Medical Making''~\cite{Lakshmi_poc} and ``DIY Assistive Technology''~\cite{buehler2015sharing} personalized fabrication by nontechnical experts is becoming an emerging and critical domain.
TactStyle can be employed to customize assistive devices with specific textures, enhancing grip, comfort, or usability tailored to the unique needs of the user. Figure~\ref{fig:application_examples}e illustrates this by applying a `rough rock' texture to the handle of a walking stick. This texture generates a rough heightfield, which post-fabrication, significantly increases surface friction, thereby improving grip and stability for the user. Such tactile enhancements are particularly valuable for assistive devices, where safety and ease of use are critical, offering a practical solution that can be tailored to the specific requirements of individuals with mobility challenges.

%\maxine{Do we have an example of this?
 %\subsection{Architectural and Interior Design}
 %Architects and interior designers could use this technology to generate tactile textures for walls, floors, and furniture, allowing clients to feel surface materials in virtual or physical models before finalizing design choices.}
\section{Discussion and Future Work}
% \faraz{\hl{highlight the limitations.}}
TactStyle demonstrates an ability to replicate both visual and tactile features from an image input, allowing creators to stylize their 3D models for both accurate color replication, and expected tactile properties. In this section, we discuss TactStyle's current limitations, and its possible extensions in the future. 

\faraz{
\subsection{Opportunities for Richer Datasets}
TactStyle’s performance and robustness to a diversity of textures is dependent the quality and diversity of its training dataset. Currently, the model utilizes the CGAxis repository~\cite{cgaxis_pbr_20_parquets}, which provides 500 texture-heightfield pairs across five material categories: \textit{Parquets}, \textit{Wood}, \textit{Rocks}, \textit{Walls}, and \textit{Roofs}. The high quality of the available images, and their corresponding heightfields allowed us to train the image-generation model with a high accuracy. While this dataset offers a diverse selection of real-world textures, this does not cover all types of textures encountered by humans. Additional material categories such as fabrics, metals, and organic surfaces could enhance the model’s generalizability, and allow personalization of tactile surfaces in fashion, automotive design, etc. Moreover, expanding the dataset to include dynamic material properties, such as elasticity, thermal responsiveness, or friction, could enable TactStyle to model textures with more complex interactions, further improving the reproducibility of their tactile properties.  
}

\subsection{Cross-Modal Texture Design}
TactStyle is able to replicate specific tactile descriptors from both expected tactile features extracted from the visual texture, and the perceived tactile properties correlated to the original heightfield. This combined replication of expected and perceived tactile properties allows for a cross-model design of textures. Recent work in VR and Haptics~\cite{degraen2021weirding} have explored novel ways to design and map user-defined tactile properties in virtual reality, such as voice. TactStyle approaches a similar problem, but in the fabrication domain, allowing users to apply textures that have `expected' tactile properties. In the future, this approach can be extended to text prompts, allowing users to describe their expected tactile response, and fabricate a 3D model with such tactile properties.  

\subsection{Incorporating Material Properties of Textures}

% \faraz{mimic textures of material properties - different printing materials, metamaterials, controlling post-processing techniques.}

The material properties of textures play a critical role in defining their tactile experience. Hardness and Scratchiness, for instance, are closely related to the rigidity and resistance of a material, directly affecting how a surface feels when touched. Currently, TactStyle operates by taking images as input to generate tactile features. While this method effectively aligns visual and tactile experiences, there is potential to enhance its accuracy by incorporating material descriptors as input. ~\faraz{In this study, we standardized the material used for all texture samples for consistency, to evaluate the accuracy of generated heightfields in replicating tactile perception. However, tactile perception is also influenced by material-specific properties such as compliance, thermal conductivity, and surface friction~\cite{degraen2021capturing}. Future work could explore how these properties can be replicated by predicting material types that approximate their tactile properties. Additionally, integrating novel approaches like metamaterials~\cite{ion2016metamaterial} could provide new avenues for tailoring and enhancing texture replication across diverse applications.} 

% In the future, TactStyle could be extended with teto process both visual and material property descriptors, enabling the system to learn the relationships between these properties and tactile sensations. By integrating these descriptors, TactStyle could create more accurate texture heightfields, improving its ability to replicate textures not just from visual perception but based on a deeper understanding of underlying material properties.

\subsection{Analyzing Visuo-Haptic Properties Together}
TactStyle currently evaluates visual and tactile perceptions separately, identifying key differences between expected tactile properties based on visual cues and the actual tactile perceptions of textures. These findings highlight an interplay between visual and haptic modalities in shaping texture perception~\cite{skedung2013feeling}. Future work could explore this property of tactile perception, and leverage visuo-haptic mismatches to create novel experiences, such as ``\textit{impossible materials}'' that visually appear soft but feel rigid, defying conventional expectations. Additionally, photochromic materials have been used in prior work to create re-programmable multi-color surfaces. Such materials offer opportunities to dynamically link visual and tactile feedback to create novel dynamic textures.

\subsection{3D Model Generation with accurate texture information} 
Recent Generative AI methods have enabled users to generate novel 3D models from scratch based on image and text prompts~\cite{xu2024instantmesh, gao2022get3d}. However, while current systems excel in generating visual representations of textures, they often lack the capacity to accurately generate the tactile properties on these materials. Since TactStyle works with image modality as well, an extension of TactStyle could allow creators to provide an image of description of a novel object and its expected tactile properties, allowing creators to not only create novel digital artifacts but also fabricatable designs with accurate texture information. By extending generative tools to encode material properties, these models could also propose materials to fabricate the object such that the tactile experience is closely approximated. 

\section{Conclusion}

In this paper, we present TactStyle, a system that allows users to stylize 3D models using image prompts, replicating both visual appearance and tactile properties. By extending generative AI techniques, TactStyle generates tactile features as a heightfield and applies them to 3D models. A quantitative study demonstrates significant improvements over traditional stylization methods. In a psychophysical experiment with 15 participants, we evaluate TactStyle's ability to create textures perceived as similar to both visually expected tactile properties and the original texture's tactile features.

Our findings show that TactStyle successfully aligns visual and tactile properties, enabling more realistic 3D model personalization. This work opens up new possibilities in cross-modal design, and future work can expand TactStyle by incorporating material descriptors to further enhance its tactile accuracy.

% In this paper, we introduce TactStyle, a novel system that allows users to stylize 3D models based on image prompts to replicate not only the aesthetic appearance but also the tactile properties. TactStyle extends traditional generative AI techniques for image generation to generate tactile properties of 3D models as a heightfield, and apply it to 3D models. We conducted a quantitative study that shows statistically significant improvement over traditional stylization approaches. In a psychophysical experiment with 15 participants, we evaluate TactStyle's ability to create textures that are perceived to be similar to both visual expected tactile properties, and the texture's original tactile properties. Our results demonstrate that TactStyle effectively bridges the gap between visual and tactile design, producing textures that align closely with user expectations based on visual cues while preserving essential tactile characteristics. This work opens up new possibilities in cross-modal design, enabling creators to seamlessly combine visual and tactile elements for more immersive and realistic 3D model creation. Future work will focus on expanding TactStyle's capabilities by incorporating material descriptors, further enhancing its ability to generate accurate tactile experiences from a wide range of inputs.

\begin{acks}
\changes{We would like to extend our sincere gratitude to the MIT-Google Program for Computing Innovation for their generous support, which made this research possible. Furthermore, we thank Varun Jampani from Stability AI, Vrushank Phadnis, Federico Tombari, and Fabian Manhardt from Google Research, and Martin Nisser from University of Washington for their valuable insights and feedback on this research.}
\end{acks}

% \begin{acks}
% \changes{We would like to extend our sincere gratitude to the MIT-Google Program for Computing Innovation for their generous support, which made this research possible. Furthermore, we thank Varun Jampani, Yingtao Tian, Vrushank Phadnis, Yuanzhen Li, and Douglas Eck from Google for their valuable insights and feedback on this research.}
% \end{acks}

% \balance
\bibliographystyle{ACM-Reference-Format}
\bibliography{references}

\appendix
\appendix

\changes{
\section{Perception Study Assessment Readings}
\label{sec:appendix_perception}

\subsection{Comparing Visual and Tactile Readings}
\label{sec:appendix_perception_diff}
\subsubsection{Hardness}
The ratings of hardness significantly differed depending on the presented texture (${\chi}^2(3) = 29.398$, $p < 0.001$). For the Original samples, significant differences were found when compared to the Visual ($W = 2476.5$, $p < 0.001$) and Stylization samples ($W = 1161.5$, $p < 0.01$), but no significant difference was observed with the TactStyle samples ($W = 927.5$, $p > 0.05$). For the Visual samples, significant differences were found with TactStyle ($W = 2978.5$, $p < 0.001$) and Stylization samples ($W = 4319.5$, $p < 0.001$). Additionally, the TactStyle samples differed significantly from Stylization samples ($W = 1080.5$, $p < 0.01$).

\subsubsection{Roughness}
The ratings of roughness differed significantly depending on the presented texture (${\chi}^2(3) = 46.48$, $p < 0.001$). For the Original samples, significant differences were found with Visual ($W = 5768.0$, $p < 0.001$), TactStyle ($W = 4511.0$, $p < 0.001$) and Stylization samples ($W = 4347.5$, $p < 0.001$). For the Visual samples, significant differences were also found with Stylization samples ($W = 6526.0$, $p < 0.01$), but no significant difference with the TactStyle samples ($W = 8503.0$, $p > 0.05$). The TactStyle samples differed significantly from the Stylization samples ($W = 7110.0$, $p < 0.01$).

\subsubsection{Bumpiness}
The ratings of bumpiness significantly differed depending on the presented texture (${\chi}^2(3) = 95.72$, $p < 0.001$). For the Original samples, no significant difference was observed with the Visual samples ($W = 7070.0$, $p > 0.05$), but significant differences were found with both the TactStyle ($W = 6257.5$, $p < 0.01$) and Stylization  ($W = 2296.0$, $p < 0.001$). For the Visual samples, no significant difference was observed with the TactStyle  ($W = 8242.5$, $p > 0.05$), but a significant difference was found with the Stylization  ($W = 3901.5$, $p < 0.001$). Additionally, the TactStyle samples significantly differed from the Stylization samples ($W = 3326.0$, $p < 0.001$).

\subsubsection{Scratchiness}
The ratings of scratchiness significantly differed depending on the presented texture (${\chi}^2(3) = 45.20$, $p < 0.001$). For the Original samples, significant differences were found when compared to both the Visual ($W = 6383.5$, $p < 0.001$) and Stylization samples ($W = 4940.5$, $p < 0.001$), but no significant difference was observed with the TactStyle samples ($W = 6907.0$, $p > 0.05$). For the Visual samples, significant differences were found with both the TactStyle ($W = 6379.0$, $p < 0.01$) and Stylization samples ($W = 8198.0$, $p < 0.05$). Additionally, the TactStyle samples significantly differed from the Stylization samples ($W = 4379.0$, $p < 0.001$).

\subsubsection{Stickiness}
The ratings of stickiness significantly differed depending on the presented texture (${\chi}^2(3) = 24.53$, $p < 0.001$). For the Original samples, significant differences were observed with both the TactStyle ($W = 3726.0$, $p < 0.01$) and Stylization samples ($W = 3685.5$, $p < 0.001$), but no significant difference was found compared to the Visual samples ($W = 5656.5$, $p > 0.05$). For the Visual samples, no significant difference was found with the TactStyle samples ($W = 6985.0$, $p > 0.05$), but a significant difference was found with the Stylization samples ($W = 5679.0$, $p < 0.01$). Additionally, no significant difference was found between TactStyle and Stylization samples ($W = 5316.0$, $p > 0.05$).

\subsubsection{Uniformity}
The ratings of uniformity significantly differed depending on the presented texture (${\chi}^2(3) = 25.02$, $p < 0.001$). For the Original samples, no significant difference was observed compared to the Visual ($W = 7504.0$, $p > 0.05$), TactStyle ($W = 6783.0$, $p > 0.05$), or Stylization samples ($W = 7393.5$, $p > 0.05$). For the Visual samples, significant differences were found with both the TactStyle ($W = 5534.0$, $p < 0.001$) and Stylization samples ($W = 7579.0$, $p < 0.05$). Additionally, no significant difference was observed between the TactStyle and Stylization samples ($W = 8565.5$, $p > 0.05$).

\subsubsection{Isotropy}
The ratings of isotropy significantly differed depending on the presented texture (${\chi}^2(3) = 27.13$, $p < 0.001$). For the Original samples, no significant difference was observed compared to the Visual ($W = 7527.5$, $p > 0.05$) or TactStyle samples ($W = 7359.5$, $p > 0.05$), but a significant difference was found with the Stylization samples ($W = 6651.5$, $p < 0.01$). For the Visual samples, no significant difference was found with the TactStyle samples ($W = 7825.0$, $p > 0.05$), but a significant difference was observed with the Stylization samples ($W = 5726.0$, $p < 0.001$). For TactStyle samples, a significant difference was found with the Stylization samples ($W = 5085.5$, $p < 0.001$).

\subsection{Perceptual Correlations}
\label{sec:appendix_perception_corr}
\subsubsection{Hardness}
The ratings of hardness were found to significantly correlate depending on the presented texture. For the Original samples, significant correlations were found with Visual samples ($r = 0.28$, $p < 0.001$), the TactStyle samples ($r = 0.64$, $p < 0.001$), and the Stylization samples ($r = 0.47$, $p < 0.001$). For the Visual samples, significant correlations were found with the TactStyle samples ($r = 0.28$, $p < 0.001$) and the Stylization samples ($r = 0.25$, $p < 0.001$). Finally, for the TactStyle samples, significant correlations were found with the Stylization samples ($r = 0.61$, $p < 0.001$).

\subsubsection{Roughness}
The ratings of roughness were found to significantly correlate depending on the presented texture. For the Original samples, significant correlations were found with the Visual samples ($r = 0.22$, $p < 0.01$) and the TactStyle samples ($r = 0.28$, $p < 0.001$), but no significant correlation was observed with the Stylization samples ($r = 0.08$, $p > 0.05$). For the Visual samples, a significant correlation was found with the TactStyle samples ($r = 0.33$, $p < 0.001$), but no significant correlation was observed with the Stylization samples ($r = -0.023757$, $p > 0.05$). Finally, for the TactStyle samples, no significant correlation was observed with the Stylization samples ($r = 0.055$, $p > 0.05$).

\subsubsection{Bumpiness}

The ratings of bumpiness were found to significantly correlate depending on the presented texture. For the Original samples, significant correlations were found with the Visual samples ($r = 0.31$, $p < 0.001$) and the TactStyle samples ($r = 0.23$, $p < 0.01$), but no significant correlation was observed with the Stylization samples ($r = 0.16$, $p > 0.05$). For the Visual samples, a significant correlation was found with TactStyle samples ($r = 0.17$, $p < 0.05$), but not with Stylization samples ($r = -0.021$, $p > 0.05$). Finally, for the TactStyle samples, no significant correlation was observed with the Stylization samples ($r = 0.14$, $p > 0.05$).

\subsubsection{Scratchiness}

The ratings of scratchiness were found to significantly correlate depending on the presented texture. For the Original samples, a significant correlation was found with the Visual samples ($r = 0.20$, $p < 0.05$) and the TactStyle samples ($r = 0.29$, $p < 0.001$), but no significant correlation was observed with the Stylization samples ($r = 0.03$, $p > 0.05$). For the Visual samples, no significant correlation was observed with the TactStyle samples ($r = 0.10$, $p > 0.05$) or the Stylization samples ($r = -0.18$, $p > 0.05$). Finally, for the TactStyle samples, no significant correlation was found with the Stylization samples ($r = 0.17$, $p > 0.05$).

\subsubsection{Stickiness}

The ratings of stickiness were found to significantly correlate depending on the presented texture. For the Original samples, significant correlations were found with the Visual samples ($r = 0.23$, $p < 0.01$) and the TactStyle samples ($r = 0.31$, $p < 0.001$), but no significant correlation was observed with the Stylization samples ($r = 0.11$, $p > 0.05$). For the Visual samples, significant correlations were found with both the TactStyle samples ($r = 0.26$, $p < 0.001$) and the Stylization samples ($r = 0.21$, $p < 0.01$). Finally, for the TactStyle samples, a significant correlation was found with the Stylization samples ($r = 0.29$, $p < 0.001$).

\subsubsection{Uniformity}
For the Original samples, a significant correlation was observed with the Visual samples ($r = 0.20$, $p < 0.05$), and with TactStyle samples ($r = 0.074$, $p < 0.05$), but not the Stylization samples ($r = 0.072$, $p > 0.05$). For the Visual samples, a significant correlation was found with the TactStyle samples ($r = 0.20$, $p < 0.05$), but no significant correlation was observed with the Stylization samples ($r = -0.10$, $p > 0.05$). Finally, for the TactStyle samples, no significant correlation was found with the Stylization samples ($r = 0.04$, $p > 0.05$).

\subsubsection{Isotropy}

The ratings of isotropy were found to significantly correlate depending on the presented texture. For the Original samples, a significant correlation was observed with the TactStyle samples ($r = 0.29$, $p < 0.001$), and with the Visual samples ($r = 0.15$, $p < 0.05$), but not with the Stylization samples ($r = 0.13$, $p > 0.05$). For the Visual samples, a significant correlation was found with the TactStyle samples ($r = 0.19$, $p < 0.05$), but not with the Stylization samples ($r = 0.04$, $p > 0.05$). Finally, for the TactStyle samples, no significant correlation was found with the Stylization samples ($r = 0.16$, $p > 0.05$).

}
\end{document}